\documentclass[paper=A4,11pt]{scrartcl}
\usepackage[bookmarks=false]{hyperref}
\usepackage{epsfig}
\usepackage{graphicx}
\usepackage{hyperref}
\usepackage{epstopdf}
\newcommand{\eq}[1]{\begin{equation}#1\end{equation}}

\newcommand{\subs}[1]{_\mathrm{#1}}

 \newcommand{\mpl}{M\subs{p}}

\newcommand{\be}{\begin{equation}}
\newcommand{\ee}{\end{equation}}
\newcommand{\bea}{\begin{eqnarray}}
\newcommand{\eea}{\end{eqnarray}}

\newcommand\sigb{\bar\sigma}
\newcommand\ytt{\tilde {y}}

\newcommand\mzt{{\tilde{m}}_0}
\newcommand\Azt{\tilde{A_0}}

\newcommand{\nnu}{\nonumber\\}

\newcommand{\oot}{\overline {126}}
  \usepackage{amsmath}
  \numberwithin{equation}{section}
 
\begin{document}
 \vfil

 \vspace{3.5 cm}
\title{\Large{\bf  { Supersymmetric Seesaw Inflation }}}
 \author{Charanjit S. Aulakh and Ila Garg  }
%\date{}
\maketitle

\normalsize\baselineskip=15pt

 {\centerline {{\it
Dept. of Physics, Panjab University}} {\centerline{ \it
{Chandigarh, India }}}}

 {\centerline {\rm E-mail: aulakh@pu.ac.in }}

\vspace{1.5 cm}

\large {\centerline{\bf {ABSTRACT }}}
\normalsize\baselineskip=15pt

\vspace{1. cm} Supersymmetric Unified theories   which incorporate
a renormalizable Type I seesaw mechanism for small   neutrino
masses  can also   provide slow roll inflection point inflation
along a flat direction associated with a   gauge invariant
combination of the Higgs, slepton and right handed sneutrino
superfields. Inflationary parameters are related to the Majorana
and Dirac couplings responsible for neutrino masses with the scale
of inflation set by  a right-handed neutrino mass $M_{\nu^c} \sim
10^6-10^{12}$ GeV. Tuning of the neutrino Dirac and Majorana
superpotential  couplings and soft Susy breaking parameters is
required to enforce flatness of the inflationary potential. In
contrast to previous inflection point inflation models the cubic
term is dominantly derived from superpotential couplings rather
than soft A-terms.  Thus  since  $M_{\nu^c}>>M_{Susy}$ the tuning
condition is  almost independent of the soft supersymmetry
breaking parameters and therefore  more stable. The required fine
tuning  is also less stringent than  for Minimal SUSY Standard
Model (MSSM) inflation or Dirac neutrino ``A-term'' inflation
scenarios due to the much larger value of the inflaton mass.
Reheating proceeds via `instant preheating'  which rapidly dumps
all the inflaton energy into a MSSM mode radiation bath  giving a
high reheat temperature
 $T_{rh} \approx M_{\nu^c}^{\frac{3}{4}}\, 10^{6}$ GeV $\sim 10^{11}- 10^{15} $ GeV.
   Thus our scenario  requires  large
gravitino mass  $> 50 $ TeV to avoid a gravitino problem. The
`instant preheating' and Higgs component of the inflaton also
imply a `non-thermal' contribution to  Leptogenesis  due to
facilitated production of right handed neutrinos during inflaton
decay.  We  derive the tuning conditions for the scenario to work
in the realistic New Minimal Supersymmetric SO(10) GUT  and show
that they can be satisfied by realistic fits.
\normalsize\baselineskip=15pt

\newpage
\section{Introduction}
Primordial inflation is perhaps the simplest dynamical mechanism
which can explain the seed fluctuations\cite{primordstruct} for
the cosmic microwave background (CMB) radiation, and thus for the
formation of  large scale structures. Although a large number of
inflationary models exist in the literature the
  majority of them are not grounded in any realistic model of particle physics, thus leaving them unconstrained by
  anything beyond the few parameters so far gleaned from measurements~\cite{COBE,WMAP7} of the Cosmic Microwave background.
 Models where inflation is  driven not by a generic  scalar field  but by an  inflaton
 intimately tied\cite{MSSMflat} to the Standard Model gauge group and spectrum  carry an obvious appeal.
Moreover in order to have a successful (and calculable)
reheating into the Standard Model degrees of freedom as is
 required for the success of Big Bang Nucleosynthesis,
the inflaton must carry definite Standard Model gauge and Yukawa charges so that the inflaton condensate
can efficiently decay  into SM degrees of freedom after the end of inflation.

The suggestion ~\cite{MSSMflat} that inflation can be naturally
embedded within the Minimal supersymmetric (SUSY) Standard Model
(MSSM), with generic gravity mediated (i.e $N=1$
%MARK
supergravity type : but we assume  canonical K$\ddot{a}$hler
potential ) soft supersymmetry breaking terms, is an attractive
scenario which enables us to connect the microscopic origin of
inflation to cosmological evolution on the  largest scales. Models
of this type are typically based on slow roll inflation associated
with ``flat directions''  in the MSSM field space (along which the
D-term potential vanishes). A well known theorem \cite{holoflat}
allows one to use holomorphic gauge invariants formed from
 chiral superfields as coordinates for the D-flat manifold of the scalar field space of SUSY gauge theories. The  flat
directions
%MARK
  are lifted by supergravity generated soft supersymmetry breaking terms and by  non
  renormalizable terms in the MSSM effective superpotential.
Such models ( also called ``A-Term Inflation'' models
\cite{akm,hotchmaz,lythdimo}) typically  require a fine tuning
between the soft terms to ensure an inflection or saddle point of
the field potential where the vacuum energy density drives a burst
of inflation but nevertheless allows ``graceful exit'' due to the
absence of a local minimum and the  associated potential barrier
which would prevent exit.   In such models the (usually
non-renormalizable)   terms that lift D-flatness of the inflaton
potential  are  hypothesized rather than deduced
 from a well defined underlying renormalizable model.
  Thus while they answer some of the relevant issues they have much scope for improvement.
 One may consider how to deduce
 the effective non-renormalizable superpotential by integrating out heavy
  fields from an  underlying theory, or one may look for
 minimal extensions of the MSSM which may (like inflationary GUTs)
    support inflation even at the renormalizable level.

The first definite signal of physics beyond the SM came from
neutrino oscillations which are now accepted as evidence of non
zero neutrino masses in the milli-eV range. However the nature of
neutrino masses,  i.e whether they are  of Dirac or Majorana type,
is still unsettled. In the first case light neutrino masses are
understood as being the consequence of highly suppressed
 Yukawa  couplings, ($ {\cal L}= y_\nu {\overline N_R} H_u L  +.... ; \, y_\nu\sim m_\nu/M_W\sim 10^{-12}$ )
  7 or more orders of magnitude smaller than the charged fermion Yukawa couplings. To be dominantly of Dirac type
  these masses should be accompanied by highly suppressed right handed
  Majorana neutrino masses, $ M_{\nu^c}   \sim 0.1 eV$ or less. Conversely one may
  generate    small  effective (Type I seesaw\cite{seesaw}) neutrino Majorana  masses ($m_\nu\sim (m_{\nu}^D)^2/M_{\nu^c}$)
  for the left handed neutrinos if the right handed neutrino  masses $M_{\nu^c}$ take the
  large values permitted by their vanishing SM gauge charges.  In this case the Dirac masses
   of the neutrinos need not be suppressed by
  ultra small Yukawa couplings as required in  the Dirac mass case.

In \cite{akm,hotchmaz} an intriguing connection was   made between
the smallness of the (Dirac) neutrino masses and  flatness of the
inflaton potential within  the MSSM extended by the addition of
$U(1)_{B-L}$ gauge group and right handed neutrinos. The inflaton
field was a gauge invariant $D$-flat direction, $N H_{u}L$, where
$N$ is the right handed sneutrino, $H_{u}$ is the MSSM Higgs which
gives masses to the up-type quarks, and $L$ is the slepton field.
The gauge invariant superpotential term $y_{\nu} N H_{u}L $
generates the tiny (Dirac) neutrino masses  due to the
aforementioned tiny neutrino Yukawa coupling  ($y_\nu\sim
10^{-12}$). When coupled with   soft trilinear and bilinear
supersymmetry breaking terms of mass scale $\sim 100 ~GeV$  to
$10~  TeV $  the associated {\it{renormalizable}} inflaton
potential can then be fine tuned to achieve inflection point
inflation consistent with Wilkinson Microwave Anisotropy
Probe(WMAP) 7 year data\cite{akm,hotchmaz}.

Since the seesaw\cite{seesaw} explanation for neutrino masses is
arguably preferable to the ad-hoc small Dirac masses explanation
it is natural to ask if it too supports  inflation.  \emph{Prima
facie} such a scenario could face   obstacles in meeting the
requirements of the neutrino-inflaton  scenario i.e  ultra small
superpotential couplings, and TeV scale  trilinear/mass terms.
 Generic Type I seesaw relies upon large right handed neutrino
Majorana masses which are generated by breaking of $B-L$ symmetry
by vevs $V_{B-L}>> 10^{10} GeV $. An inflaton  involving the right
handed sneutrino  will then have (supersymmetric)  mass
contributions as large as the righthanded neutrino mass. The
cogency of the seesaw lies in not artificially singling out
neutrino Yukawas to be ultra small. With large $V_{B-L}$ the Dirac
coupling of the neutrino need not be suppressed by hand. Indeed
for, normal hierarchy, one obtains the third generation light
neutrino masses $m_{\nu}\sim y_3^2 v_{EW}^2/M_{\nu^c}\sim 0.1 eV$
for $y_3\sim 1, M_{\nu^c} \sim 10^{15} $ GeV. Such large couplings
and masses would completely destroy the needed flatness of the
inflationary potential. However reflection shows that a negative
conclusion may be unwarranted since at least three generations of
neutrinos and their superpartners are in play. So  there is
considerable scope for much smaller superpotential couplings: the
neutrino Yukawa coupling eigenvalues  could have the typical
values associated with up type fermions   while off diagonal
components matched   the tiny Majorana couplings in smallness. Off
diagonal flat directions ($N_A H L_B, A\neq B$=1,2,3) can serve
just as well as diagonal ones , in fact we shall see they are
required in the realistic New Minimal SO(10) GUT implementation of
our scenario .

  Furthermore the popular Leptogenesis\cite{leptogen} scenario
strongly hints at right handed neutrino masses in the range
$10^{6} $ to $10^{12}$ GeV. So for $V_{B-L}\sim M_X> 10^{16}$ GeV
the superpotential couplings $f_A, A=1,2,3$ (we will work in a
basis where these couplings are diagonal), which generate right
handed neutrino masses $M_{\nu^c_A}\sim f_A V_{B-L}$, are very
small ($ f_{A}\sim 10^{-9}$ to $10^{-4})$. Thus  the required
ingredients for an inflaton in the Type I seesaw scenario are
already   present. Note that since generic Type I seesaw requires
that  $B-L$ is   broken at a high scale, issues concerning the
efficient decay of the conjugate sneutrino component of the
inflaton  via their Yukawa couplings will need to be addressed.
Reheating in our scenario  proceeds via the so called `instant
pre-heating'  mechanism\cite{feldkoflinde} resulting in a high
reheat temperature due to rapid dump of the inflaton energy into
MSSM modes. The Higgs component of the inflaton
implies\cite{ahnkolb} a non-thermal contribution to leptogenesis.

Issues regarding natural values for superpotential couplings come
into focus when viewed in the context of the so called Minimal
Left Right supersymmetric models\cite{MSLRMs} and their embedding
in GUT models\cite{rparso10,MSGUTs}. SUSY Left-Right Models are
advantaged due to their protection of R-parity as a gauged
discrete symmetry, which provides a stable lightest supersymmetric
particle (LSP). They  simultaneously and naturally implement
Seesaw mechanisms for neutrino masses\cite{MSLRMs}. Moreover such
models have also been incorporated in the realistic and predictive
New Minimal Susy SO(10)  grand unified
theories(NMSGUT)\cite{NMSGUTs,nmsgut3} where all the hard
parameters of the MSSM are fitted in terms of fundamental
parameters of the GUT and soft SUSY breaking parameters (of the
Non-Universal Higgs masses (NUHM) type) defined at the Unification
scale $M_X\sim 10^{16}- 10^{18} $ GeV. Such GUTs have viable Bino
dark matter candidates and make distinctive predictions for the
type of SUSY spectra observable at the LHC. In 2008, well before
 the discovery of Higgs mass of around 125 GeV
in 2011-2012  and the consequent realization that a general
framework such as the  phenomenological MSSM (pMSSM) requires that
the soft trilinear couplings $A_{t,b}$ be \emph{large}, we
concluded\cite{NMSGUTs} that \emph{the NMSGUT would  be falsified
} by its failure  to fit the down type quark masses \emph{unless}
$A_0,\mu$ were in the 10's of TeV : leading to a mostly decoupled
superspectrum with only the LSP, gauginos and possibly a light
slepton in the sub-TeV range ! The experimental data has now
forced this realization on practitioners of MSSM
parametrology\cite{djouadi}. In the NMSGUT it was a pre-diction.
In the NMSGUT the
 successful fitting of fermion masses necessarily entails
ultrasmall  neutrino Majorana-Yukawa couplings leading to (first
generation) right handed neutrino Majorana masses as small as
$10^6$ GeV. Taken together with the possibility of small values
for the light generation Yukawa Dirac  couplings it is possible to
implement viable inflection point inflation by suitable tuning
 at the supersymmetric level itself. This is technically more appealing
than a tuning applied to soft susy parameters which, being
unprotected by SUSY, are unstable. We derive the tuning conditions
for the NMSO(10)GUT and show how to satisfy them explicitly.

In Section $\bf{2}$ we review and summarize the generic
renormalizable single scalar inflaton inflection point model and
calculate its slow-roll parameters, power spectrum and spectral
index so as to use these  results    with the supersymmetric
%MARK
Type I Seesaw model(SIMSSM), with supergravity soft terms, once we
have shown that it generates a suitable potential of the
renormalizable type. In Section $\bf{3}$ we see how a generic
Supersymmetric $SU(3)\times SU(2)_L\times U(1)_R\times U(1)_{B-L}$
model, with generic supergravity type soft supersymmetry breaking
terms, which contains the essentials of the Type I Susy seesaw
implemented in the Minimal Susy LR Models and in R-parity
preserving Susy GUTs,  provides an attractive Inflationary
scenario in parallel with its achievement of realistic neutrino
masses. In Section $\bf{4}$ we discuss the general features of
reheating in this model and remark on the types of Leptogenesis
that can arise from the inflaton and right handed neutrino decay.
In Section $\bf{5}$ we give a discussion of the embedding in the
NMSGUT and the inflationary parameters associated with realistic
fits. We conclude with a brief discussion.

\section{Generic  Renormalizable Inflection point inflation }

A generic renormalizable  inflection point inflation model can be
formulated in terms of a single
 complex field $\varphi$. Such a model\cite{akm,hotchmaz}   can reproduce the
observed\cite{WMAP7} inflationary  power spectrum $P_R=(2.43\pm
0.11)\times 10^{-9}$, spectral index $n_s=.967\pm 0.014$ and scale
invariance $ k dn_s/dk \simeq 0 $. It is also relevant to note
that the ratio of tensor  to scalar power spectrum
  amplitudes $r= \frac{P_T}{P_R}$ is known to be less than about
  0.5. After extremizing with respect
 to the angular degree of freedom (which has positive curvature
 and cannot support inflation) one is left with the  potential for
a real  degree of freedom $\phi$ in  the complex scalar
 inflaton field $\varphi$
\bea V= {\frac{h^2}{12}} \phi^4  -  {\frac{A h}{6\sqrt{3}}} \phi^3
  + {\frac{M^2}{ 2}}\phi^2 \eea
Here $A,h, \phi$ are real and  positive without loss of
generality. The formulae we derive in this section are applicable
to any single inflaton theory with a renormalizable potential.

In the model of \cite{akm} the $A,M$ receive dominant
contributions from trilinear and quadratic  soft supersymmetry
breaking parameters: $A,M\sim 10^2-10^4 ~$ GeV. A very small
neutrino Yukawa coupling $\sim 10^{-12}$ and a high degree of fine
tuning between A and M    is necessary to reproduce the observed
inflation parameters\cite{hotchmaz,lythdimo}. In our work however
the contributions from soft supersymmetry breaking terms play a
negligible role. The controlling mass scale is much higher , the
required size of the yukawa couplings is larger and the degree of
fine tuning is much less.

 It is convenient to trade the parameter
A for a fine-tuning parameter $\Delta$ by replacing $A=4 M{\sqrt{
1-\Delta }}$ ($\Delta=\beta^2/4$ in the notation of
\cite{hotchmaz}). The inflection point at  \bea \phi_0
=\frac{\sqrt{3} M}{h}(1-\Delta +O(\Delta^2)) \qquad :\qquad
V''(\phi_0)=0\eea is also a saddle point ($V'(\phi_0)=0$) when
$\Delta =0$. For small $\Delta$  \bea V(\phi_0)&=&V_0=\frac{M^4}{4
h^2} (1 + 4 \Delta)\qquad;\qquad
V'(\phi_0)=\alpha=\frac{\sqrt{3} M^3 \Delta}{ h} \nonumber\\
V'''(\phi_0)&=&\gamma=\frac{2 M h}{\sqrt{3}}(1-2
\Delta)\label{leadingV}\eea  If the coupling $h $ is tiny $V_0\ >>
M^4$ and   $\phi_0>>M/h$. Notice that $\gamma$ tends to be quite
 small due to the smallness of $h$, while $\alpha$ is small
 (but non-zero\cite{lythdimo}) because it is tuned to be small.
The large vacuum energy and flatness of the potential around
$\phi_0$ then imply that if $\phi$ starts with a value close to
$\phi=\phi_0 $ and a small field velocity the universe will
execute slow roll inflation as the field $\phi$ rolls slowly down
through a narrow field interval of width $\Delta \phi \sim
V_0/\gamma M_p^2$ below $\phi_0$. Around the inflection point
$\phi_0$, we can   write  the inflection point inflation potential
in the form \bea
V(\phi)=V_0+\alpha(\phi-\phi_0)+\frac{\gamma}{6}(\phi-\phi_0)^3
+{\frac{h^2}{12}}(\phi-\phi_0)^4   \label{eq-Vinf-gen} \eea The
last term is  essentially negligible since $h^2 $ is very small by
assumption.\\
 The slow roll parameters are defined as($ M_{p }= 2.43 \times 10^{18} \,GeV$)  \bea
 \eta(\phi)&=&\frac{M_p^2 V''}{V} \simeq {\frac{M_p^2}{V_0}}\gamma (\phi-\phi_0)\nonumber
\\ \epsilon(\phi)&=&\frac{M_p^2}{2} (\frac{V'}{V})^2 \simeq(\alpha
+\frac{\gamma}{2}(\phi-\phi_0)^2)^2({\frac{ M_p^2}{2V_0^2}}) \nonumber
 \\  \xi&=&\frac{M_p^4V' V'''}{V^2} \simeq \frac{M_p^4 \alpha \gamma}{V_0^2}
  \label{slrlprm}\eea

  The small first and third Taylor coefficients
$\alpha,\gamma$  determine\cite{lythdimo,lythstew,liddleleach} the
measured parameters of inflation ($P_R,n_s$) once the field values
($\phi_{CMB},\phi_{end}$) at the time of horizon  entry of the
``pivot'' momentum scale ($k_{pivot}=0.002$ Mpc${}^{-1}$) and at
termination of the slow roll are
fixed\cite{lythdimo,liddleleach}(on the basis of an overall
cosmogonic scenario and the consistency of the slow roll
approximation ($\eta(\phi_{end})\approx 1$) respectively).
$k_{pivot}$ corresponds to a representative scale of current
cosmological observations. The field value at the beginning of
inflation is of notional interest only. It is the number
($N_{CMB}=N(\phi_{CMB})$) of e-folds of inflation left to occur
after field value $\phi_{CMB}$  reached at the time when the
fluctuation scale of interest($k_{pivot}$) left the comoving
horizon ( i.e $k=a_k H_k$) during inflation that is of
significance. This number is determined by the overall history of
the Universe from primordial times\cite{liddleleach}. Plausible
inflationary cosmogonies   require $ 40<N_{CMB} < 60$ and this
severely restricts the inflation exponents.

The field value at the end of slow roll inflation $\phi_{end}$ is
defined as the value where \bea \eta(\phi_{end})\simeq
\frac{\gamma(\phi_{end}-\phi_0) M_p^2}{V(\phi_0)}\simeq 1  \eea
which gives \bea\phi_0-\phi_{end}=\frac{V_0}{\gamma
M_p^2},\label{phiend}\eea Then in the slow roll approx
$\ddot\phi\simeq 0 , \dot\phi=-V'(\phi)/3 H$,
 where $H=\sqrt{V(\phi_0)/(3 M_p^2)}$ is the (constant) inflation
  rate during slow roll inflation, one has
   \bea N(\phi)&=& -3 \int_{\phi}^{\phi_{end}}\frac{H^2}{V'(\phi)}
   d\phi \nonumber \\
    &=& \sqrt{\frac{2}{ \alpha\gamma}}\frac{V_0}{M_p^2}
    \big(\arctan\sqrt{\frac{\gamma}{2\alpha}}(\phi_0-\phi_{end})-
    \arctan\sqrt{\frac{\gamma}{2\alpha}}(\phi_0-\phi )\big)\label{Nphi}\eea
and conversely

\bea  \phi(N)&=&{\frac{\phi_{end} +\phi_0 (\phi_0-\phi_{end})
    \sqrt{\frac{\gamma}{ 2\alpha}}
     \tan\sqrt{\frac{\alpha\gamma}{2}}\frac{N M_p^2}{V_0}+
     \sqrt{\frac{ 2\alpha}{\gamma}}
     \tan\sqrt{\frac{\alpha\gamma}{2}}\frac{N M_p^2}{V_0}}
     { 1+(\phi_0-\phi_{end})
    \sqrt{\frac{\gamma}{ 2\alpha}}
     \tan\sqrt{\frac{\alpha\gamma}{2}}\frac{N M_p^2}{V_0}}}
    \eea

It is worth remarking that this inversion of the function
$N(\phi)$  was derived without assuming that $\phi_{end}<<\phi(N)$
\cite{lythdimo}. Together with an interpolating function derived
below it allows us to obtain analytic formulae for the  relations
required among the parameters of the inflationary potential for
successful inflation: avoiding tedious and opaque  graphical
methods \cite{hotchmaz}.

 The observed Cosmic Microwave Background(CMB) data \cite{WMAP7} pose constraints on the power
spectrum and spectral index for modes around the pivot scale.
Barring non-standard scenarios where the post-inflationary period
is punctuated by episodes of modified expansion,   the number of
remaining e-folds at the time the pivot scale left the horizon
during inflation may be estimated by using the standard Big Bang
thermal cosmogony along with estimates of the reheating behaviour
of the universe after inflation. This gives\cite{liddleleach}
\eq{N\subs{pivot} =
 65.5+ \ln\frac{\rho_{rh}^{\frac{1}{12}}V_0^{\frac{1}{6}}}{\mpl } }
where $\rho_{rh}$ is the energy density after reheating and $V_0$
the potential value during inflation. The reheating behaviour of
the $NLH$ flat direction inflaton is quite different from the the
original quadratic chaotic inflation
 models for conjugate sneutrino inflation\cite{murayana,elliyana}.
    The tripartite  composition of the inflaton out of $L,H,\tilde \nu^c$ degrees of freedom will
    ensure that the bulk of the energy in the inflaton will be dumped into light degrees of freedom on
    the very first oscillation. For the present we merely assume that the reheating is immediate
         so that one can set $\rho_{rh}=V_0$ in $N_{Pivot}$. We then find that for M in the range
$10^{6} - 10^{12}\, GeV$, $h$ lies between $10^{-9.5}\,$ to
         $10^{-6.5}\,$ and then  $N_{pivot}=N_{CMB}=51\pm 5$ adequately covers the
         possible range. Even if this range is lowered by effects
         of reheating or non-standard cosmogonies the effect on
         the relevant exponents will prove to be marginal.
   Although the observed CMB is actually a
combined spectrum of modes exiting the horizon around
$|N-N\subs{pivot}|\leq 5$, we can approximate and regard it as the
single spectrum from the mode that exits the co-moving  horizon
when $ \phi=\phi_{CMB}$ only. Thus $\phi=\phi_{CMB}$ is the field
value near $\phi_0$ where the inflation giving rise to observable effects today kicks in
(when $N_{CMB}$ e-folds of inflation are remaining).
 The power spectrum and spectral index we see today are then
$P_R(\phi(N\subs{CMB}))$ and $n_s(\phi(N\subs{CMB}))$
respectively, where $|N_{CMB}-N\subs{pivot}|< 5$.

The slow roll inflation formula for the power spectrum of the mode
that is leaving horizon when the inflaton rolls to $\phi$
is(\cite{lythstew}) \eq{P_R(\phi)=\frac{ V_0}{24 \pi^2
\mpl^4\epsilon(\phi)},\label{eq-inf-PR}} and the corresponding
spectral index and it's variation with momentum is
 \bea n_s(\phi)&\equiv&
 1+2\eta(\phi)-6\epsilon(\phi) \label{eq-inf-ns}\nonumber \\
  {{\cal D}_k}(n_s)&=&\frac{kdn_s(\phi)}{dk} =-16\epsilon\eta + 24
\epsilon^2 + 2 \xi^2\eea The ratio of tensor to scalar
perturbations $r=\frac{P_T}{P_R}=16\epsilon.$  In practice
$\epsilon,\xi$ are so small in the narrow region near $\phi_0$
where slow-roll inflation occurs that their contribution to $n_s$
is negligible. Thus ${\cal D}_k(n_s)$ is negligible i.e. the
spectral index is scale invariant in the observed range, as is
allowed by observation so far.

To search for sets of potential parameters
 $M,h,\Delta$ compatible with $P_R,n_S,N_{CMB} $  in their
 allowed ranges one may proceed as follows. First one uses the
 chosen (within experimental range)
  values of $P_R,n_s$  and given $M,h$ to
   define
  \bea \epsilon_{CMB}&=&\frac {V_0}{24 \pi^2 M_p^4 P_R}\nonumber \\
  \eta_{CMB} &=& \frac{(n_s-1)}{2}\label{epsetacmb}\eea

  From  $\epsilon_{CMB},\eta_{CMB}$ one may deduce
  $\alpha_{CMB},\phi_{CMB}$    using the
 eqns.(\ref{slrlprm}) \bea\phi_{CMB}&=&\phi_0 +
{\frac{V_0 \eta_{CMB}}{ \gamma  M_p^2}} \nonumber\\
\alpha_{CMB}&=& \sqrt{2  \epsilon_{CMB}} {\frac{V_0}{M_p}}
-{\frac{V_0^2\eta_{CMB}^2}{2 \gamma  M_p^4} } \eea

 The required fine-tuning $\Delta$  is then  \be \Delta= {\frac{h\alpha_{CMB}}{\sqrt{3}M^3}}=(\frac{M}{4 h M_p})^4
 (\frac{16 h^2 M_p}{3 \pi M {\sqrt{P_R}}}-(1-n_s)^2)\ee
 $\alpha_{CMB},\Delta$ should emerge real and positive
 and using $\{\alpha_{CMB},\phi_{CMB}\}$ in the formula for
$N_{CMB}$ one should obtain a sensible value  in the range
$N_{CMB}=51\pm 5$. Positivity of $\Delta$ (a local minimum
develops if  $\Delta$ is negative leading to eternal inflation)
requires \be h^2 \geq M \frac{3 \pi(1-n_s)^2\sqrt{P_R}}{16 M_P}\ee
Using eqns.(\ref{leadingV},\ref{phiend}\ref{epsetacmb}) in
eqn(\ref{Nphi}) we have \bea N_{CMB}&=& \frac{1}{z} \arctan\frac{2
z (1+ n_s)}{8 z^2 + 1 - n_s}\label{NCMB}\\z&=& (\frac{h^2 M_p}{3
\pi M\sqrt{P_R}} - \frac{(1-n_s)^2}{16})^{\frac{1}{2}} \label{z}\eea By
solving eqn.(\ref{NCMB}) for $z=z_0(N_{CMB},n_s)$ one obtains the
general relation between $h$ and $M$ : \bea \frac{h^2}{M}&=&
\frac{3 \pi \sqrt{P_{R}}}{M_{P}}
(z_0^2(N_{CMB},n_s)+\frac{(1-n_s)^2}{16})\eea and then
\bea\Delta&=& \frac{16 M^2 z_0^2(N_{CMB},n_s) }{9 \pi^2 M_P^2 P_R
((1-n_s)^2+ 16 z_0^2(N_{CMB},n_s))^2  }\eea Where
$z_0(N_{CMB},n_s)$ is the solution of eqn (\ref{NCMB}). An
excellent approximation to the the required function in the region
of interest in the $N_{CMB},n_s$ plane is given by the Taylor
series around $n_s^0=0.967,N_{C}^0=50.006$ :\bea
&z_0(N_{CMB},n_s)=.0238
-0.0006 (N_{CMB}-N_C^0)+ 0.2407 (n_s- n_s^0) \nonumber\\
&+ 0.000022 \frac{(N_{CMB}-N_C^0)^2}{2}
 -3.70875 \frac{(n_s-n_s^0)^2}{2}+ 0.002353 (N_{CMB}-N_C^0)(n_s-
n_s^0)\nonumber\\& -.0000015 \frac{(N_{CMB}-N_C^0)^3}{6}+
 8.79982 \frac{(n_s-n_s^0)^3}{6}
 -0.000788 \frac{(N_{CMB}-N_C^0)^2 (n_s-n_s^0)}{2}\nonumber\\
& -0.7536 \frac{(N_{CMB}-N_C^0) (n_s-n_s^0)^2}{2}\eea

\begin{figure}
\centering
\includegraphics[scale=0.8]{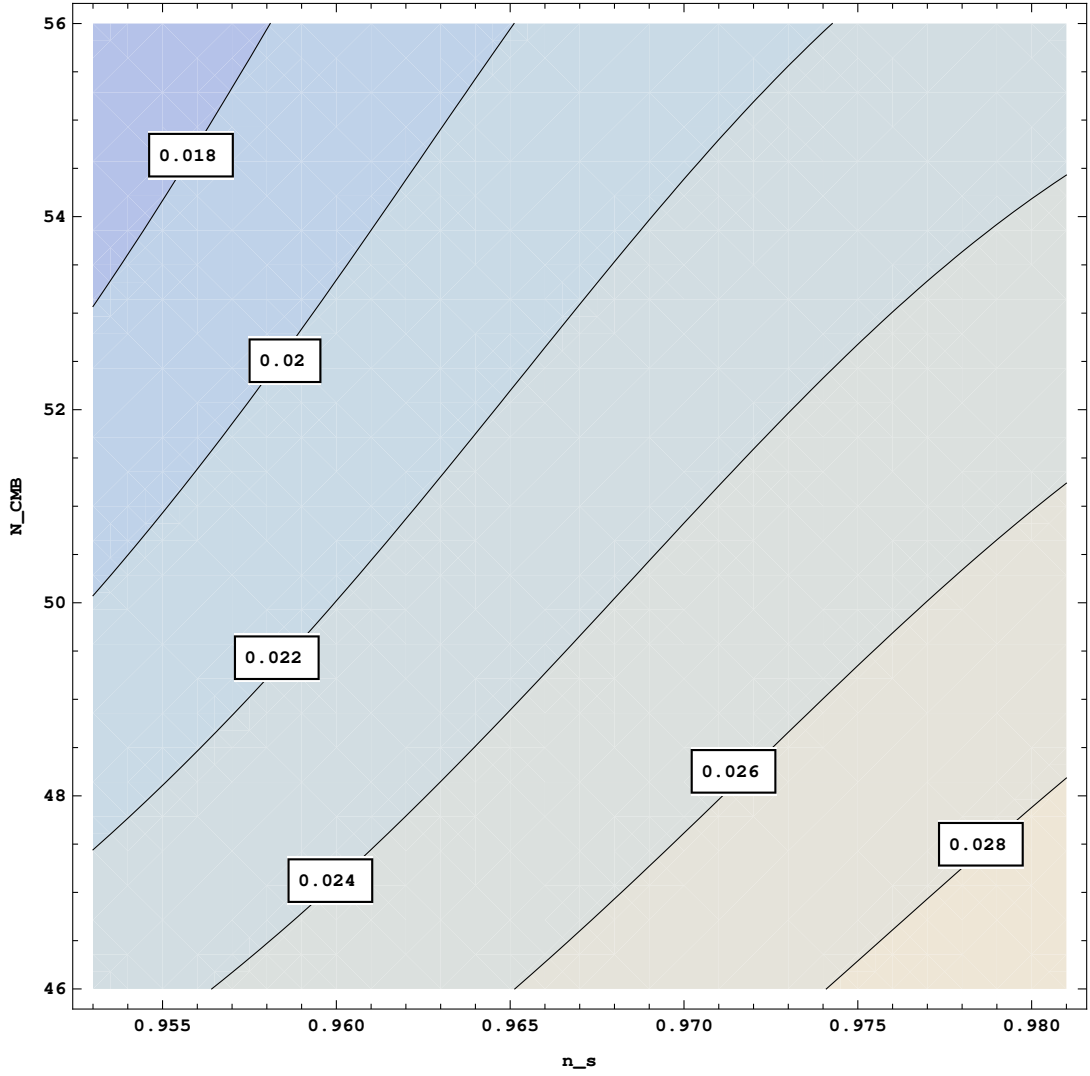}
\caption{$z_0 $ contours in the $N_{CMB},n_s$ plane. The variation
shown contributes to the small range of permitted magnitudes   for
$h^2/M, \Delta/M^2$ etc}
\end{figure}

 In Fig. 1 we have plotted the contours
of $z_0(N_{CMB},n_s)$ in the $N_{CMB},n_s$ plane and one sees that
the variation of $z_0$ is rather modest. So for the plausible
range $46<N_{CNB}<56$ one obtains a tight constraint on the
exponents in the relation between h,$\Delta$ and M: \be h^2 \sim
10^{-24.95 \pm 0.17} (\frac{M}{GeV})\,\,\,\,;\,\,\,\ \Delta \sim
10^{-28.17 \pm .13}(\frac{M}{GeV})^2\label{hDeltavsM}\ee
\begin{figure}
\includegraphics[scale=0.8]{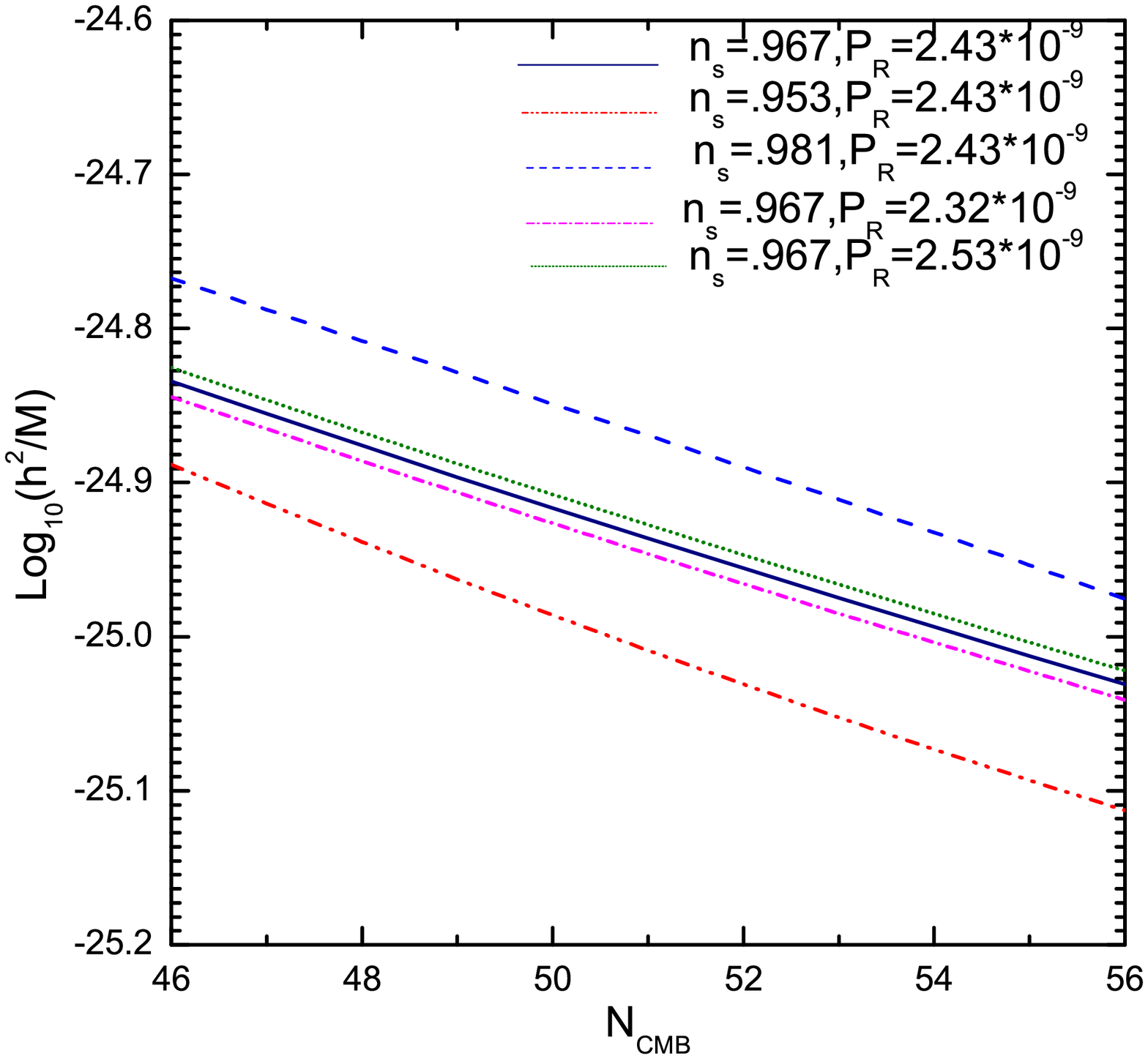}
\caption{Variation of exponent of $\frac{h^2}{M}$ with $N_{CMB}$
for different values of $n_s,P_R$.}
\end{figure}
\begin{figure}
\includegraphics[scale=0.8]{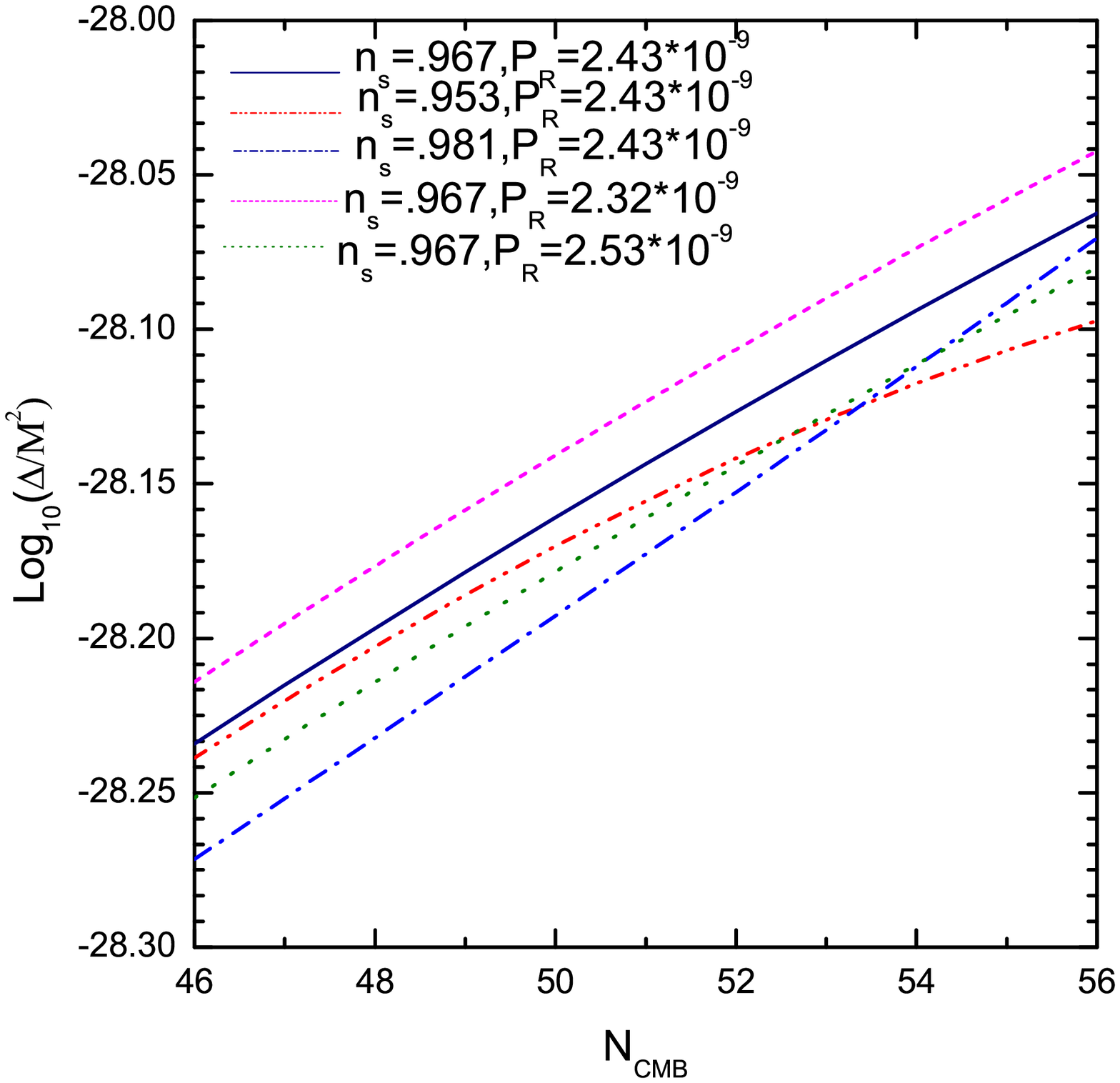}
\caption{Variation of exponent of $\frac{\Delta}{M^2}$ with
$N_{CMB}$ for different values of $n_s,P_R$.}
\end{figure}

 We have estimated the maximum variations
in the exponents corresponding to the quoted errors in the WMAP 7-
year data \cite{WMAP7} from the graphs in  Fig. 2 and  Fig. 3.

However a clearer qualitative understanding results from noticing
that for $N_{CMB} \sim 50$, $Z_{0}\approx \frac{1.2}{N_{CMB}}$
solves eqn.(\ref{NCMB}) to a good approximation. Then
eqn.(\ref{z}) gives \be \frac{h^2}{M} \approx \frac{3 \pi}{M_{P}}
\frac{\sqrt{P_R}}{N_{CMB}^2} \approx \frac{2.75 \times
{10^{-22}}}{N_{CMB}^2} \approx 10^{-25}\ee \be \frac{\Delta}{M^2}
\approx \frac{4.14 \times 10^{-34}}{N_{CMB}^2 P_R} \approx
10^{-28.2} GeV^{-2}\ee Thus these simple approximate   expressions
give   effectively the same results  as the more carefully derived
expressions in eqn.(\ref{hDeltavsM}).
 Thus we have viable inflation with \be V_0
\sim \frac{M^4}{h^2}\sim (M)^3\times 10^{25} \, GeV \sim
10^{43}-10^{61} \, GeV^4\ee \be H_0 \sim \sqrt{\frac{V_0}{M_P^2}}
\sim 10^{3 }-10^{12} \, GeV \,\,\,\,;\,\,\,T_{max} \sim
V_0^{\frac{1}{4}} \sim 10^{11}-10^{15} \,GeV \ee It is clear from
eqn.(\ref{hDeltavsM}) that the fine-tuning measure grows   with
$M$ so that $\beta=\sqrt{\Delta}$ can be as large as $10^{-2}$ for
$M\sim 10^{12}$ GeV. Due to the large value of the inflaton mass
compared to the case of MSSM inflation\cite{MSSMflat} or Dirac
neutrino inflation\cite{akm,lythdimo}  the   fine-tuning of
parameters required is much less severe  and no additional
dynamics need be invoked to make it
plausible\cite{hotchmaz,enqmazstep}. It is also important to note
that the ratio  of tensor to scalar perturbations $r\simeq
16\epsilon \simeq 2 (M/10^{14}GeV)^3$. Since $M$ is at most the
heaviest right handed neutrino mass $\sim 10^{12}$ GeV it is clear
that it is difficult to get $r > 10^{-6}$. Thus measurement of
tensor perturbations via the Cosmic Microwave background
polarization at the $r\sim 10^{-3}$ level or larger would not be
compatible with inflection point inflation controlled by the right
handed neutrino mass. Any renormalizable single inflaton model
must respect these generic constraints and yield values of its
associated parameters that are sensible in terms of the other
(particle) physics that it describes. It remains to specify the
Renormalizable Susy seesaw Inflaton scenario and consider the
NMSGUT as a self contained realistic test bed.

\section{  Supersymmetric   seesaw Inflaton model }

The essentials of the Supersymmetric seesaw inflation scenario
 may be captured by considering a
model with gauge group $ SU(3)\times SU(2)\times U(1)_R\times
U(1)_{B-L} $ and the field content of the MSSM with some
additional superfields. Soft supersymmetry breaking terms are of
the supergravity type [i.e trilinears proportional to yukawa
couplings and universal, or universal except for Higgs (NUHM
scenario), soft scalar masses]. The essential fields beyond the
MSSM consist of a right handed Neutrino chiral multiplet   $N
[1,1,-1/2,1]$ and a field ${ S}[1,1,1,-2]$ whose vev generates the
large Majorana masses $M_\nu$ ($10^6-10^{14}$ GeV) for the
conjugate neutrinos $\nu^c_A \equiv N_A$ via a renormalizable
superpotential coupling $3 \sqrt{2} f_{AB} S \nu^c_A \nu^c_B   $.
  Additional fields $\Theta_i$ which serve to fix
 the vev of $S$ are also present as in Minimal Supersymmetric
 Left Right Models (MSLRMs)\cite{MSLRMs} and in GUTs that embed them
\cite{rparso10,MSGUTs,NMSGUTs}. The other essential component of
the scenario is neutrino  Dirac mass   generating Yukawa couplings
$y_{AB}, A,B=1,2,3$ in the superpotential.  These couple the right
handed neutrinos to the Left chiral lepton doublets
$L_A=\begin{pmatrix}\nu & e \end{pmatrix}_A^T, A=1,2,3$. $L_A$
transform  as $L[1,2,0,-1]$ and the up type  Higgs doublet type
field as $H[1,2,1/2,0]$ so that $y_{AB}N L_A H_B$ is a gauge
invariant term in the Superpotential. Of course each such doublet
present in the underlying theory must have its complementary
doublet transforming as e.g. $\overline{H}$[1,2,-1/2,0] to cancel
anomalies. The relevant flat direction is assumed to extend out of
the minimum of the supersymmetric potential corresponding to the
breaking of the gauge group down to the MSSM symmetry  \bea
SU(3)\times SU(2)\times U(1)_R\times U(1)_{B-L} \rightarrow
SU(3)\times SU(2)\times U(1)_Y\eea This leads to a Type I seesaw
plus MSSM (SIMSSM) effective theory.

After the breaking one has $Y= 2 T_{3R} + (B-L)$ where $T_{3R}$ is
the $U(1)_R$ generator. Note that unlike the case of the Dirac
neutrino masses scenario \cite{akm} $B-L$ is \emph{not} a gauge symmetry down to low
energies. This can have important consequences for nucleosynthesis
and matter domination since the heavy right handed neutrinos must
find a non-gauge channel to decay through. In the present case
this channel must perforce be a Yukawa coupling since the
right-handed neutrinos are singlets of the low energy (SM) gauge
group. This is in contrast to the Dirac scenario where a low scale
of B-L breaking is assumed so that $\nu^c$ can decay via gauge
couplings.

The fields $\tilde N$(the chosen conjugate sneutrino ), $\tilde
\nu $( chosen left sneutrino flavour from a Lepton doublet L with
suitable Yukawa couplings) and the light neutral  Higgs field
(from the doublet $H$ with $Y=+1$) may be parametrized in terms of
the flat-direction associated with the gauge invariant $NLH$ as
\begin{equation}
\tilde{N}= \tilde{\nu}=h_0=\frac{\varphi}{\sqrt{3}}=\phi
e^{i\theta};~~~\phi\geq 0,~~ \theta\in[0,2 \pi)\, \label{inflaton}
\end{equation}
The additional fields $\Omega_i$, unspecified at the moment, are
assumed to be coupled to $S$  in such a way that extremization of
the SUSY potential using $F_{\Omega_i}=0,~~D_\alpha|_{\phi=0}=0 $
fixes the vev of S: $<S>=\bar\sigma/\sqrt{2}$ without constraining
the inflaton field $\varphi$. This is of course true in the
Minimal Susy LR models\cite{MSLRMs} and renormalizable  Susy
SO(10) GUTs \cite{MSGUTs,NMSGUTs} which are our inspiration.

The vanishing of the  $D$-term  for the $B-L$ generator requires
$\Omega_i$ to include the companion field(s) ${\overline{S}}
[1,1,-1, 2]$ which have a vev of equal magnitude as $S$ in order
to preserve SUSY through the symmetry breaking down to the MSSM
symmetry at high scales. This is just as in MSLRMs and R-parity
preserving GUTs \cite{MSLRMs,rparso10,MSGUTs,NMSGUTs}.
 The gauge invariance of $NLH$ ensures that the $D$-terms for
 the flat direction vanish. Thus at scales $\phi\sim \bar\sigma>>M_{S}$
 where SUSY is exact the relevant superpotential is given by:
\begin{equation}
    W=    3\sqrt{3} y N   \nu h+ 3 f{\sqrt{2}} S NN +...= y\varphi^3 +f{\sqrt{2}}
     S  \varphi^2 +...
\label{WSIMSSM}\end{equation}
where $ h,f,\bar\sigma$ can be taken real without loss of generality.
The right handed neutrino Majorana mass will be $M_\nu=6f\sigb $.

Since  the equations of motion of the unperturbed vacuum imply
$<F_S>=0,\\ <S>= \bar\sigma/\sqrt{2}$ this superpotential leads to a
flat direction potential
\begin{eqnarray}
    V_{susy}&=&|3 y \varphi^2 + 2 f\bar\sigma\varphi|^2 +   2 |f
    \varphi^2|^2\nonumber\\
    &=& f^2     \left[ (2+9{\tilde{y}}^2){\phi}^4 +12
    {\tilde{y}}{\phi}^3 \sigb\cos\theta + 4 \sigb^2{\phi}^2\right]
\end{eqnarray}
Here $\tilde{y} =y/f$  and we see that $\sigb$ sets the mass
scale. Minimizing with respect to $\theta$ gives $\theta=\pi$. In
so far as we are here interested only in the inflationary dynamics
(once parameters have been tuned to ensure an inflection point in
the plateau region where $|\varphi| \sim {\bar\sigma}$)  we can
focus on just the real part of $\varphi$ and set
 $\varphi=-\phi$ with $\phi$ real and positive near the inflection
 point but free to fall into the well around $\phi=0$  and
 oscillate around that value. The imaginary part $\phi'$ of
 $\varphi$ has a large curvature $V_{\phi'\phi'}\sim
 {\bar\sigma}^2 $ in the plateau region. Since
 $V_{\phi'}|_{\phi'=0}=0$ it is consistent to consider  the dynamics in the
 real $\varphi $ plane alone as a leading approximation. The
 effect of jitter in the $\phi'$ direction when the dynamics is
 initiated with $\phi'\neq 0$ can be studied numerically as a
 correction to the dynamics of the inflaton field $\phi$.

  In addition one also expects a contribution to the potential
  from the $\mu $ term for the Higgs doublets together with
SUSY breaking quadratic and cubic soft terms, which we assume to
be of the type generated by supergravity, but with non universal
Higgs masses,  i.e of the form:
\begin{eqnarray}
    V_{soft} &=& \big [A_0 (y \varphi^3  + f \sqrt{2} S\varphi^2)+ h.c\big ] +m_{\tilde f}^2
    \sum_{\tilde f} | {\tilde f}|^2 + m_{ H}^2 |  H|^2  + m_{\bar H}^2 |\bar H|^2  \nnu
    &=& f^2    \left[\ytt\Azt\phi^3 \sigb \cos{3\theta} +  \Azt \sigb^2{ \phi}^2
    \cos{2\theta}+\mzt^2 \sigb^2 { \phi}^2 \right]
    \end{eqnarray}
here $\mzt=m_0/{f\sigb}, \Azt= 2 A_0/{f\sigb} $. The soft mass
 $m_0$ receives contributions from the sfermion and Higgs soft
masses as well as the $\mu$ term \bea m_0^2= \frac{2 m_{\tilde
f}^2 + \overline{m}_H^2}{3}\eea $m_{\tilde f,H}$ are the sfermion
and up type Higgs soft effective masses at the unification scale
($\overline{m}_H^2=m_H^2+|\mu|^2$). Since these masses and $A_0$
should be in the range $10^2-10^5$ GeV while the righthanded
neutrino masses lie in the range $10^{6}-10^{12}$ GeV, it is clear
that $\tilde m_0,\tilde A_0 $ are   small parameters and even for
the large values of $m_0, A_0\sim 10^5$ GeV found in the NMSGUT
$\mzt,\Azt <<1 $.  Thus these terms cannot significantly change
$\theta=\pi$ assumed earlier. The total inflaton potential is then
\eq{V_{tot}=f^2\left((2+9\tilde y^2) \phi^4-(\tilde A_0+12)\tilde
y\bar\sigma\phi^3 +  (\tilde A_0+\tilde
m_0^2+4)\bar\sigma^2\phi^2\right).\label{eq-inf-Vtot}} Thus we
have  a generic quartic inflaton potential of the same type as in
Section $\bf{2}$ but  the parameter values  in the case of Type I
seesaw are quite different from the light Dirac neutrino case.  We
  have the  identification
of parameters \bea h&=&f\sqrt{12(2 + 9 \ytt^2)}\nonumber\\
A&=&\frac{3 f (\Azt +12)\ytt \sigb}{\sqrt{(2 + 9
\ytt^2)}}\nonumber\\
M^2&=& 2 f^2 \sigb^2(4+\Azt +\mzt^2)\nonumber \\
\Delta&=& (1-\frac{A^2}{16 M^2})\nonumber\\
&=&\left(1-\frac{9\tilde y^2(\tilde A_0+12)^2}{32(2+9\tilde
y^2)(\tilde A_0+\tilde m_0^2+4)}\right)\label{paramident}\eea For
seesaw models the natural magnitude for the neutrino Dirac mass
is, $m_{\nu}^D
>1 MeV $ \hfil\break
(i.e $\, |y_\nu^D| > 10^{-5}$ and then the limit $m_{\nu}<<0.01
eV$ for the lightest neutrino (assuming direct hierarchy) implies
$M_{\nu^c} > 10^6$ GeV). Since the preferred values for the Susy
breaking scale are smaller than 100  TeV (at most) it follows that
the maximum value of $|\Azt|,|\mzt| \sim 0.1 $ and they could be
much smaller for more typical larger values of the conjugate
neutrino masses $M_{\nu^c} \sim 10^8 $ to $ 10^{12}$ GeV.  It is
then clear from the corresponding range $\Delta\sim 10^{-12} $ to
$10^{-4}$ that the coupling ratio $\ytt=y/f$ becomes ever closer
to exactly $\ytt =4/3$ as M increases and even for $M\sim 10^6$
GeV differs from $1.333 $ only at the second decimal place. Thus
to a good approximation $ h=6{\sqrt{6}} f $. Then it follows from
  the  Eqs.(\ref{hDeltavsM})and (\ref{paramident}) that \bea  f
  &\simeq &
10^{-26.83 \pm 0.17}(\frac{\overline{\sigma}}{GeV})\qquad;\qquad
 M \simeq 10^{-25.38 \pm 0.17}(\frac{\overline{\sigma}}{GeV})^2\nonumber\\
 \Delta &\simeq & 10^{-78.93 \pm 0.47} (\frac{\overline{\sigma}}{GeV})^4\eea
 The range $M\sim
10^{6.6}$ to $10^{10.6}$ GeV  corresponds nicely to  $10^{16}
~GeV< \sigb < 10^{18} ~GeV $:  as is natural in single scale Susy
SO(10) GUTs\cite{rparso10,MSGUTs,NMSGUTs,nmsgut3}.  $f$ increases
with $\sigb$  with values below to $10^{-11}$ achievable in the
NMSGUT only with difficulty.  Of course in MSLRMs, since there are
no GUT constraints on $\sigb$, one can assume somewhat wider
ranges for these parameters.

 In all relevant cases $\Delta < 10^{ -4}$ is required. Thus the above equations imply
that ${\tilde y}^2$ must be close to the value \be {{\tilde
y}_0}^2 = \frac{64}{9}  {\frac {4+\Azt +\mzt^2}{16-8\Azt-32
\mzt^2+\Azt^2}}\ee
  Here $\Azt,\mzt \sim O(M_S/M_{\nu^c})
<<1$,  hence $\tilde y_0$ is rather close to $4/3$ and the
equality is very close for larger $M\sim f\sigb$ since then
$\Azt,\mzt$ are tiny.
 This then is the type of fine tuning that supports the development of
inflation in SIMSSM models. We see that the measure of
severity of fine tuning $\beta=\sqrt{\Delta} \sim 10^{-2} -
10^{-6}$ compares quite  favourably with the case of the MSSM or
Dirac neutrino inflaton since there $\beta\sim 10^{-12} $ to
$10^{-10}$ due to the low values of the inflaton mass in those
cases. The dominant component of the fine tuning in the present
case  is a fine-tuning of superpotential parameters, which is
radiatively stable due to non renormalization theorems. Specially
for large $\sigb > 10^{16} ~GeV$ the Type I Susy seesaw can
provide a rather attractive inflationary seesaw with a natural
explanation for neutrino masses and   weaker tuning demands on the
radiatively unstable Susy breaking parameters than the extreme and
unstable fine-tunings demanded by typical inflection point
scenarios and in particular the Dirac neutrino model \cite{akm}. Moreover,
unlike the chaotic sneutrino inflaton
scenario\cite{murayana,elliyana}, no trans-Planckian vevs are
invoked.

\section{Reheating and Leptogenesis}

After inflation concludes the energy stored in the inflaton will
be transferred into a thermal bath of the  MSSM degrees of
freedom. Determination of the time required to thermalize the
inflaton energy and the resulting    reheat  temperature $T_{rh}$
(i.e the maximum temperature of the thermal bath after
thermalization) requires understanding the post-inflationary
dynamics of the $LHN$ flat direction inflaton. An important issue
that can be tackled at the level of the effective SIMSSM is
generation of the the cosmological baryon number
asymmetry($n_B/n_\gamma$) via Leptogenesis\cite{leptogen}.
Although a detailed analysis of these issues requires a separate
publication, the existence of previous detailed studies  of
preheating\cite{feldkoflinde} in a MSSM flat direction inflaton
model\cite{alfermaz} and of non-thermal Leptogenesis in a
preheating model\cite{ahnkolb}  make the generalizations required
to combine the two ideas in the context of supersymmetric seesaw
inflation easy to outline, but too long to derive, here.
Supersymmetric seesaw inflation offers an attractive synthesis
fulfilling the need expressed in \cite{ahnkolb} :

 "\emph{There
have been many models of leptogenesis. A hallmark of our model is
the economy of fields. The only undiscovered fields are the
inflaton, $\phi$, the standard model Higgs, h, and the
right-handed neutrino, N. There are very good reasons for
suspecting that all exist! The only unfamiliar aspect of our model
is the strong coupling of the inflaton field to the Higgs field.
While there is no reason to preclude such a coupling, it would be
very interesting to find particle-physics models with a motivation
for the coupling. }"

 In our model the the inflaton is
itself partly comprised of the Higgs field and therefore fulfills
the requirements of \cite{ahnkolb} exactly, besides bringing
together a number  of other related  streams of thought. We remark
however that the situation is made more complex by the high
reheating temperature associated with the large inflaton mass.
Thus both thermal and non-thermal leptogenesis may contribute to
the generation  of $n_B/n_\gamma$.

Due to the gauge(H,L)  and third generation yukawa($H$) coupled
components of the inflaton the inflaton energy is likely to decay
very rapidly (i.e in decay time $\tau_{dec} << H_{infl}^{-1} \sim
(h M_p)/M^2$) through the so called `instant preheating''
mechanism\cite{feldkoflinde,ahnkolb,alfermaz}. In this mechanism
the preheating dynamics results in a rapid decay (well within a
Hubble time)  of the complete inflaton vacuum energy into a
radiation bath which therefore  thermalizes to a temperature
determined essentially by the equality between the radiation bath
energy and the starting inflaton energy. This gives an estimate
for the
%MARK
 reheating temperature  \bea T_{rh}\sim T_{max} \sim V_0^{1/4}  \sim
M/h^{1/2}\sim 10^{11}-10^{15}\, GeV\eea The parametric dependence
is identical to  that found in \cite{alfermaz}, the difference in
scales arises only because the inflaton mass $M\sim 10^6-10^{12}$
GeV  in our model is much larger than the inflaton mass
$m_\phi\sim 0.1-10$ TeV  in \cite{alfermaz}  coming from soft
Supersymmetry breaking.

 In the preheating  mechanism  a class
(``$\chi $ type") of degrees of freedom, whose masses($m_\chi\sim g
\phi(t)$) and decay rates ($\Gamma \sim g^3 \phi(t)$) are
proportional to the instantaneous inflaton value $\phi(t)$,  are
produced non-perturbatively every time the instanton field crosses
zero. This occurs since the $\chi$ modes are ultralight for a sufficiently
large time interval around the zero crossing time during which
adiabaticity is violated ( $  {\dot\omega}_k  > \omega_k^2$ :
where $\omega_k$ is the oscillation frequency at  wave number $
k$). In our model the $\chi$ modes are the components of the
$H,L,u^c_L,u_L$ chiral superfields and the $W_\pm,B$ gauge
superfields. In fact the $\chi$ modes can be identified simply by
checking   which fields become massive in the presence of
background values of the three components of the inflaton ($\tilde
\nu,\tilde \nu^c_L,h^0$). Then with the usual superpotential (we
have suppressed generation indices)
 \be W=y^u Q_L H u^c_L  + y^d Q_L {\overline H}
  d^c_L + y^\nu L H N  + y^l L{\overline H}e^c_L +...\ee
we see that $y^u$ leads to massive $u_L,u^c_L$, $y^\nu$ leads to
massive $e_L$(one combination of the three $e_L$ ), $h^0,h^+
,\nu_L,\nu^c_L$; $y^l$ leads to massive ${\bar h}^-,e^c_L$(one
combination). Since $<H,N,L>$ preserve $U(1)_{em}$, the gauge
couplings give masses to Z (which forms a Dirac supermultiplet
with $(\nu-{\tilde h}_0)/\sqrt{2}$) and $W_{\pm} $ (form a pair of
Dirac supermultiplets with $l^{-},{h}_+$). This set  of fields are
the $\chi$ type fields whose mass varies strongly with $\phi$ as
it oscillates and whose production, when $\phi\sim 0$, and   decay
when $\phi >> M_W$, is the basis of `instant preheating'. The
inflaton vev  leaves the down quark and gluon/gluino fields and
$\bar h_0$, and some combinations of the $l^-_L,l^c_L$ fields with
light (MSSM type) masses. These light ($\psi$-type) fields will
form the first step in the decays of the $\chi$ field.
   As $<\phi>$ again increases the
$\chi$ modes become very heavy and unstable and as a result decay
rapidly(within a time $\tau_{dec}\sim \frac{h}{M g^3}<<
m_{\phi}^{-1}$)  to the light (mostly coloured) MSSM d.o.f. to
which they are coupled (dominantly via the D-terms and
gauge-yukawa terms but also via the Superpotential
 couplings for the third generation). As a result a significant
 fraction $\sim 10^{-1}$ of the inflaton condensate energy  passes into
 the light MSSM modes with every crossing resulting in complete
 transfer within $\sim 10^2$ oscillation times. \bea\tau_{osc}\sim m_{\phi}^{-1}<<
 H_{infln}^{-1} \sim (h M_p)\tau_{osc}/M \sim (1  \, - \, 150)
 \tau_{osc}\eea  Once the energy is in the light modes  MSSM
 interactions, in particular the gauge interactions,
  are sufficient to rapidly complete thermalization so
 that essentially all the inflaton energy will be thermalized
 within, at most,  a few  Hubble times after the end of inflation.
Rapid decay of the inflaton oscillation amplitude   leaves the
light modes to thermalize the energy dumped by the inflaton into a
radiation bath of all modes: which are no longer ever heavy
because the inflaton has decayed.
 The reheating temperature is \bea T_{rh} \sim ({\frac {30} {\pi^2 g_*}})^{1/4} V_0^{1/4}
 \sim T_{max} \sim 10^{11} - 10^{15} \, GeV \eea
where $g_*=228.75$ is the effective number of MSSM degrees of
freedom. The essential point is that the reheating temperature is
well above that required to produce relativistic  populations of
gravitinos : which are unacceptable if their lifetimes are  larger
than the nucleosynthesis  time $\tau_{N} \sim 1 \, sec $ since
their decay after nucleosynthesis would destroy the created
nucleons. The   straightforward and generic resolution of this
gravitino problem is if the graviton masses are sufficiently large
so that the gravitinos decay before
nucleosynthesis\cite{moroi}\bea \tau_{grav} \sim 10^5 \, sec
({\frac{1 \, TeV }{m_{3/2}}})^3 << \tau_N \sim 1 sec \eea Thus we
see that the viability of Supersymmetric seesaw Inflation strongly
indicates that the scale of supersymmetry breaking -as indicated
by the gravitino mass- should  be  above $ 50 $ TeV. The
fact\cite{NMSGUTs}  that such large supersymmetry breaking scales
are preferred by both the NMSGUT and the latest data
indicating\cite{LHC5fb} light Higgs mass $M_h\sim 125\, GeV$
rounds off  the picture nicely. Furthermore such large reheat
temperatures also ensure abundant thermal production of all
flavours of righthanded neutrinos after inflation. Their CP
violating decays into leptons can generate the   net lepton number
density which drives creation of the requires $n_B/n_\gamma$  by
Sphaleron processing\cite{leptogen}. Thus the NMSGUT can not only
accommodate inflation but is also compatible\cite{csaigck} with
(thermal) Leptogenesis \cite{leptogen} for generating the observed
baryon to entropy density  $n_B/s \sim 10^{-10}$ .

An interesting additional source of \emph{non-thermal}
leptogenesis is provided when one  realizes\cite{ahnkolb} that,
  the Higgs field H, which is one of three fields making up
the putative  inflaton direction in field space, is itself a
$\chi$ type field and is furthermore    coupled to the righthanded
neutrinos. Thus during the course of   oscillations of the field
components of the inflaton - which commence at the end of the
inflation -  the Higgs mass $m_h\sim   g_2 \phi$ fluctuates to
values both  below and above the    righthanded neutrino masses
which are   essentially constant at  $M_{\nu^c}\sim f
 {\bar\sigma} $ even in the presence of the inflaton (i.e
$\tilde\nu,h,N$) background since $g_2>>f,y$. Thus as the Higgs
mass oscillates below and above the right handed neutrino masses
one expects CP violating -therefore net lepton number producing -
inter-conversion of the Higgs with righthanded Neutrinos as in
\cite{ahnkolb}. If this net lepton number is not washed out by the
inflaton energy dump (so that a Hubble volume contains a certain
Lepton excess produced by this inter-conversion  even though
average energies are  well above the mass of the right handed
Higgs) then we may expect that a non-thermal Leptogenesis
component will add to the thermal leptogenesis due to decay of the
righthanded neutrino bath.

An important complication in the present case, that we have
glossed over in the above account, is that the $L, H$ and $N$
components of the inflaton can have quite different decay rates
once the gauge interactions are effective, since $N$ is a gauge
singlet. A proper analysis must track the evolution of all three
fields making up the inflaton -   from an initial condition(the
end of inflation) where they start out equal. This makes the
equation of motion and Boltzmann equation for the relevant degrees
of freedom significantly more complex and this requires a separate
numerical study which involves the interplay of the couplings
$f_A,y_{AB},g_2$. The  study of this evolution and the operation
of Leptogenesis in these models is now in progress.

\section{ Inflation and neutrino masses in the NMSGUT}

Finally we  consider the embedding of our generic Type I scenario
in a realistic Susy SO(10) model\cite{NMSGUTs,nmsgut3,blmdm} that
has successfully fitted the known fermion mass-mixing data and can
also be consistent with limits from B violation and other exotic
processes\cite{nmsgut3}. We will see that neutrino flavour plays a
key role in enabling inflation : the model favours an inflaton
composed of third generation conjugate sneutrino, first generation
left slepton (sneutrino) and $T_{3R}=1/2$ Higgs.

The New Minimal SO(10) GUT (NMSO(10)GUT)  uses Higgs fields in the
${\mathbf{210,126,\oot}}$ representations of SO(10) which contain
5 SM singlets whose vevs break SO(10) down to the SM gauge group
at a superheavy scale $M_X$. Three of these vevs, called
$p,\omega,a$ come from the ${\mathbf{210 }}$-plet and one each
from the ${\mathbf{ 126 (\sigma),\oot(\sigb)}}$. An explicit
solution to symmetry breaking $SO(10)\rightarrow SIMSSM $, in
terms of a simple cubic equation for a complex variable $x$ and
depending on a single parameter ratio $\xi$ was found in the third
paper in \cite{MSGUTs}. This solution preserves supersymmetry and
makes no use of the soft breaking terms which constitute a
negligible perturbation of the global susy symmetry breaking
problem\cite{weinberg}, in the sense that they   modify the
superheavy  vevs $\sim M_X\sim 10^{17}$ GeV only by terms of order
$M_S\sim 10^4 GeV$. The spectra calculated\cite{MSGUTs,NMSGUTs}
using this analytic solution  for the  the MSGUT vacuum   are the
basis of our detailed Renormalization Group analysis of grand
unification in this class of models. Questioning   the received
wisdom that large SO(10) representations make grand unification
futile\cite{dixitsher} we showed\cite{NMSGUTs,nmsgut3,ag2} that
the inclusion of threshold corrections considerably ameliorates
the problem of large gauge beta functions by allowing one to raise
the threshold corrected unification scale close to the Planck
scale and lower the gauge coupling at unification. Taken together
these features imply that even with the huge beta functions
characteristic of MSGUTs the problem of a Landau pole in the gauge
coupling may be postponed to the Planck scale : where it becomes
moot along with the structure of space time anyway.  The physics
of asymptotically strongly coupled gravity and gauge theories  is
anybody's guess (see however \cite{tas} for our speculations and
simplified model for `tamed' asymptotically strong GUTs). There
are even claims that gravity is capable of ensuring the asymptotic
freedom of any gauge theory\cite{wilrob,toms}. It is also possible
that a RG fixed surface on which the gauge coupling remains weak
in the UV may exist. In view of the many uncertainties we take the
stand that the large beta functions of the NMSGUT are  not an
issue that need prohibit the study of these minimal and realistic
theories.

 The grand unified minimum of the potential defined by the vevs
 $\Omega\equiv \{p,\omega,a,\sigma,\bar\sigma\}$ shifts only by fractions of
 order $10^{-12}$ due to supergravity mediated soft supersymmetry breaking terms.
 The D terms of SO(10) are all exactly zero for these vevs. To examine the issue of an inflaton
 corresponding to the $NLH$ flat direction in the SIMSSM  we must
 demonstrate the existence of a corresponding flat direction of the full GUT potential
 based on light(SIMSSM) field vevs. This flat direction  rolls out   of the grand unified
 minimum that defines the MSGUT vacuum with the SIMSSM as its effective theory.
  The relevant fields are the GUT scale  vev fields $\Omega$  and
 the (6) possible components $ h_i,{\bar{h}_i};i=1...6$ of the light MSSM Higgs doublet
 pair $H,{\overline H} $ together with  the  chiral lepton fields $L_A,\nu_A^c,
 A=1,2,3$.  The relevant superpotential is
 then\cite{MSGUTs,NMSGUTs}
\bea W&=& 2\sqrt{2}(h_{AB}h_1-2\sqrt{3}f_{AB}h_2-g_{AB}(h_5+i
\sqrt{3}h_6))+\bar{h}^T
{\cal{H}}(<\Omega>)h\nonumber\\&&+4\sqrt{2} f_{AB}\bar\sigma
 \bar \nu_{A} \bar\nu_B + W_\Omega(\Omega) \eea
where
  \bea
 W_\Omega(\Omega)&=& m(p^2+3a^2+6\omega^2)+2 \lambda(a^3+3p\omega^2)\nonumber\\&&+(M+\eta(p+3a-6 \omega))\sigma \bar\sigma \eea
 and \be \frac{\partial W_{\Omega}}{\partial \Omega }|_{h,\bar \nu,L = 0} =0 \,\,\,\,\,\,\,\,\,\,\,\,\,\,
  D_{\alpha}(\Omega)|_{h,\bar \nu,L = 0}=0\label{omegavac}\ee
 here  $h_{AB},g_{AB},f_{AB}$ are the yukawa coupling matrices of
the three matter 16-plets  to the $\mathbf{10,120,\oot}$ Higgs
multiplets respectively. Equation (\ref{omegavac}) defines the
MSGUT vacuum\cite{MSGUTs}.

Of the 5  diagonal D-terms of SO(10)  only those corresponding to
the generators $T_{3L},\break T_{3R},B-L $ are charge and color
neutral with vevs for $\Omega$ and $\nu,\nu^c,h_0$. The vevs
$\Omega$ do not contribute to these D terms  so their values are
 \be
 D_{3L}=\frac{g_{u}}{2}(-\sum_{i=1}^6|h_{i0}|^2+\sum_{A}|\tilde \nu_A|^2)\ee
 \be
D_{3R}=\frac{g_{u}}{2}(\sum_{i=1}^6|h_{i0}|^2-2|h_{40}|^2-\sum_{A}|\tilde
{\bar\nu}_{A}|^2)\ee \be
D_{B-L}=\sqrt{\frac{3}{8}}g_{u}(\sum_{A}(|\tilde
{\bar\nu}_{A}|-|\tilde \nu_A|^2)+2|h_{40}|^2)\ee where we have
used the fact  (paper 5 in \cite{MSGUTs} and \cite{NMSGUTs}) that
only $h_{4\alpha}=\Phi^{44}_{{\dot{2}}\alpha}$ has
$B-L=+2,T_{3R}=-1/2 $ and thus $T_{3L}=1/2$ while all others have
$T_{3R}=1/2$ and $B-L=0$. Note that $g_u$ is the SO(10) gauge
coupling in the standard unitary normalization. Thus the
D-flatness conditions are
  \be \sum_A|\tilde
\nu_A|^2=\sum_{i}|h_{i0}^2|=\sum_A |\tilde{\bar
\nu}_{A}|^2+2|h_{40}|^2 \ee For simplicity we assume that only one
generation each  of sneutrinos $\nu_A$  and conjugate sneutrinos
$\bar{\nu}_B$ contributes to the inflaton flat direction; but not
that they must belong to the same generation. At this point we
remind the reader\cite{MSGUTs,NMSGUTs} that in MSGUTs the MSSM
Higgs doublet pair is defined by fine tuning $Det({\cal{H}})\simeq
0$ so that its lightest eigenvalue $\mu \sim M_W\sim 1 $ TeV
specifies the $\mu$ term in the superpotential of the SIMSSM :
$W=\mu {\overline{H}}H+...$. The doublet pair $H,{\overline{H}}$
is a linear combination\cite{MSGUTs,ag2,ag1} of the 6 doublet
pairs of the the NMSGUT :
 \bea  h_i=U_{ij} H_j   \qquad \qquad  \bar{h}_i={\overline{U}}_{ij}
 \overline{H}_j\eea
where $U,{\overline{U}}$ are the unitary matrices that diagonalize
the doublet mass matrix ${\cal{H}}$ : ${\overline{U}}{\cal{H}} U=
Diag\{\mu,M^H_2,....,M^H_6\}$ to positive masses. To leading
approximation they can be calculated with $\mu=0=Det({\cal{H}})$.
The so called Higgs fractions :
$\alpha_i=U_{i1},\bar\alpha_i={\overline{U}}_{i1}$ , are crucial
in determining the grand unified formulae\cite{MSGUTs,NMSGUTs} for
the fermion yukawa couplings that give rise to the fermion masses.
To obtain the tree level yukawa couplings  one makes  the
replacement $ h_i,\bar{h}_i\rightarrow \alpha_i H,\bar{\alpha}_i
{\bar{H}} $ in the expressions coupling the GUT Higgs
doublets($h_i,\bar{h}_i$) to the matter fermions of the SIMSSM.
Thus in particular the neutrino Dirac coupling is ($(\tilde
h_{AB},\tilde g_{AB},\tilde
f_{AB})$=$2\sqrt{2}(h_{AB},g_{AB},f_{AB})$ ) \be
y^{\nu}_{AB}=\tilde h_{AB} \alpha_1-2\sqrt{3}\tilde
f_{AB}\alpha_2-\tilde g_{AB}(\alpha_5+i \sqrt{3}\alpha_6)\ee From
the $|F_{\bar h}|^2$ contributions to the  potential it is clear
that the involvement of any but the light Higgs doublet $H$ would
lead to GUT scale rather than conjugate neutrino scale masses for
the inflaton.  Moreover in view of the stringent upper bounds on
the fermion yukawas (see eqn.(\ref{hDeltavsM})) the involvement of
the lightest generation is unavoidable. Thus we take
$\nu_{A}=\nu_1$. However if we also take $\tilde{\bar \nu}_{
A}=\tilde{\bar \nu}_{1}$ we find that the tuning constraint has
\emph{at best} the form $|y_{11}|^2\sim 10 (|y_{21}|^2
+|y_{31}|^2)$ : which is very hard to satisfy with normal neutrino
mass  hierarchy(the case studied so far) in   (N)MSGUTs . On the
other hand with $\nu^c_A=\nu^c_3 $ there is a possibility of
satisfying the fine tuning condition. Thus our ansatz for the flat
direction fields is
 \be \tilde{\nu}_1=
\frac{\phi}{\sqrt{3}} \qquad\qquad
 h_{i0}=\frac{\alpha_{i}\phi}{\sqrt{3}}  \qquad\qquad \tilde{\bar
\nu}_{3}=\frac{\phi}{\sqrt{3}}\sqrt{1-2|\alpha_4|^2}\ee Notice the
peculiar role of the Higgs fraction $\alpha_4$ which enters the
flat direction ansatz as  as $\Gamma=1- 2|\alpha_4|^2$. As it
happens the solutions we have found earlier \cite{NMSGUTs} often
have $|\alpha_4|\sim 0.5$. Thus it is not inconceivable that
$\Gamma$ can be consistently tuned to zero by varying the GUT
parameters. The challenge is to do so without destroying the
realistic fits to the fermion data.

By varying the fields $\Omega,\nu_A,\bar{\nu}_A$ we can now easily
derive the F-term potential \bea V_{hard}&=&
\big[({y^\nu}^\dagger y^\nu)_{11} + \Gamma(|\tilde{h}_{31}|^2
 + 4|\tilde g_{31}|^2+(y^\nu{y^\nu}^\dag)_{33}) +4|\tilde  f_{33}|^2\Gamma^2 \big ] {\frac{|\phi|^4}{9}} \nonumber
 \\&&+\frac{8}{3\sqrt{3}} \tilde
 f_{33}|y^{\nu}_{31}| |\bar \sigma| {\sqrt{\Gamma}} \text
 Cos(\theta_\phi+\theta_{y^{\nu}_{31}}-\theta_{\bar \sigma})) |\phi|^3 + \\&&\big[ {\frac {|\mu|^2}{3}} +
 \frac{16}{3}|\tilde f_{33}|^2|\bar \sigma|^2\Gamma\big ]|\phi|^2   \eea
We can also  write down the generic   Supergravity(SUGRY)-NUHM
generated soft terms in terms of a common trilinear parameter
$A_0$  but
 different soft mass parameters $\tilde{m}_{\tilde
f}^2,\tilde{m}_{h_i}^2$ for the  16 plets  and the different Higgs
(we have dropped the constant term from $ M_S W(\Omega)$ assuming
it is removed by the Supergravity scenario tuning to set the GUT
scale vacuum energy to zero by tuning hidden sector parameters).
The differences among the SO(10) Higgs soft masses could be due to
renormalization from the threshold corrected unification
scale/Planck scale to the   scale $M_X^0=10^{16.25}$ GeV  at which
the SIMSSM and NMSGUT are matched in our work \cite{agk}.
\bea V_{soft}&=&A_0 W +c.c.+{\tilde{m}}_{ 16}^2|\tilde{\Psi}|^2+\sum_{i}\tilde m^2_{h_i}|h_i|^2\nonumber\\
   &=&2A_0\sqrt{\Gamma}|y^{\nu}_{31}|\frac{|\phi|^3}{3\sqrt{3}}\text Cos(3\theta_\phi+\theta_{y^{\nu}_{31}})
+\frac{4}{3}A_0 \tilde f_{33}|\bar \sigma| \Gamma |\phi|^2\text
Cos(2\theta_\phi+\theta_{\bar\sigma})\nonumber\\&&+
({\widehat{m}}_0^2 -{\frac{|\mu|^2}{3}} )|\phi|^2 \eea where \bea
{\widehat{m}}_0^2=\frac{ \tilde{m}_{16}^2}{3}(1+\Gamma)
+\sum_{i}\frac{ \tilde{m}_{h_i}^2 |\alpha_i|^2}{3}
+{\frac{|\mu|^2}{3}}
 \eea and $\tilde{m}_{16},
\tilde{m}_{h_i},A_0$ are all   $\sim O( M_S)$ Now the extreme
dominance $f_{33}|\bar\sigma|>> M_S$ implies that the phase
$\theta_\phi$ is fixed by minimizing just the term in $V_{hard}$:
\be \theta_\phi= \pi + \theta_{ \bar\sigma }
-\theta_{y^\nu_{31}}\ee We shall assume that $\theta_\phi$ is
fixed at this value.  Since the inflationary dynamics is at large
values  of $|\phi|$ and fixed $\theta_\phi$ we can work just with
a real field $\phi$. Comparing the sum of the hard and soft
potentials with the generic renormalizable inflaton potential in
Section \textbf{2}, we immediately obtain the parameter
identifications
  \bea
h&=&\frac{2}{\sqrt{3}}\big[(y^{\nu \dag} y^{\nu})_{11}+
\Gamma(|\tilde{h}_{31}|^2+4|\tilde
g_{31}|^2+(y^{\nu}y^{\nu \dag})_{33})+ 4|\tilde f_{33}|^2\Gamma^2)\big ]^{\frac{1}{2}}\nonumber\\
A&=&\frac{1}{h}(16|\tilde f_{33}|
|y^{\nu}_{31}||\bar\sigma|{\sqrt\Gamma} +
 4 |y^{\nu}_{31}|{\sqrt\Gamma}A_0 \text Cos (3
 \theta_{\bar\sigma}-2 \theta_{y^{\nu}_{31}}))\nonumber\\
M^2 &=&\frac{32}{3}|\tilde f_{33}|^2|\bar
\sigma|^2\Gamma+\frac{8}{3}A_0\tilde f_{33}|\bar \sigma|
\Gamma\text Cos (3\theta_{\bar\sigma}-2 \theta_{y^{\nu}_{31}})+2
{\widehat{m}}_0^2 \eea The fine tuning condition $A=4 M$  now
becomes \bea|y^{\nu}_{31}|^2&=& \frac{8
\Lambda_{n}}{9\Lambda_{d}-8 \Lambda_{n}(1+\Gamma)}\big
[|y^{\nu}_{11}|^2+|y^{\nu}_{21}|^2+\Gamma(|\tilde{h}_{31}|^2+4|\tilde
g_{31}|^2+|y^{\nu}_{32}|^2+|y^{\nu}_{33}|^2)\nonumber\\&&+
4|\tilde f_{33}|^2\Gamma^2\big ]\eea Where \bea
\Lambda_{n}&=&1+\frac{A_0}{4M_3}\text Cos(3\theta_{\bar\sigma}-2
\theta_{y^{\nu}_{31}})+
\frac{3{{\widehat m}_0}^2}{16 M_3^2 \Gamma}\nonumber\\
\Lambda_{d}&=& (1+\frac{A_0}{4M_3} Cos(3\theta_{\bar\sigma}-2
\theta_{y^{\nu}_{31}}))^2\eea and $M_3=\tilde f_{33}|\bar \sigma|
$ Note that in view of the ratio between the soft breaking scale
and the mass of the heaviest right handed neutrino,
$\Lambda_{n,d}$ are both very close to unity. Thus the fine tuning
condition is essentially between hard parameters  as in GUTs and
in sharp contrast to MSSM inflaton models\cite{MSSMflat}:
\bea|y^{\nu}_{31}|^2&=& \frac{8
}{1-8\Gamma}(\Gamma(|\tilde{h}_{31}|^2+4|\tilde
g_{31}|^2+|y^{\nu}_{32}|^2+|y^{\nu}_{33}|^2)\nonumber\\&&+
|y^{\nu}_{11}|^2+|y^{\nu}_{21}|^2+4|\tilde f_{33}|^2\Gamma^2)\eea
In NMSGUT fits of the fermion data we typically find  a strong
hierarchy $|y_{33}|>>|y_{32}|>>|y_{31}|>>|y_{21}|>|y_{11}|$. So it
is evident that one must tune \be \Gamma \approx 0 \,\,\,\
{\text{i.e}}\qquad\quad
 |\alpha_4|\approx {\frac{1}{\sqrt{2}}} \ee to a good accuracy. This means that the
MSSM doublet H is almost exactly 50\% derived from the doublet in
the 210 plet ! If this condition can be achieved the remaining
tuning condition is only \bea|y^{\nu}_{31}|^2&=& 8
(|y^{\nu}_{11}|^2+|y^{\nu}_{21}|^2 )\eea which is easy to enforce
in the NMSGUT.

However there is an additional demand coming from
eqn(\ref{hDeltavsM}) : $h^2/M_3 \sim ({y^\nu}^\dag y^\nu)_{11}/M_3
\sim 10^{-25} $  which is, at first glance,   much harder to
enforce. It is rather remarkable that our results in\cite{nmsgut3}
offer a quite reasonable way out of also this predicament. The
point is that \cite{nmsgut3,wright}, the yukawa couplings of
matter fermions to the MSSM Higgs receive large wave function
corrections due to the circulation of heavy fields within loops on
the lines entering the yukawa vertex. As a result the tree level
yukawa couplings of the NMSGUT must be  dressed before they can be
matched  with those in the SIMSSM :
 \bea Y_f=(1 +{\Delta}_{\bar f}^T)\cdot (Y_f)_{tree}\cdot(1+  \Delta_f)
   (1+ \Delta_{H^{\pm}} ) \eea
 Due to the large number of
heavy fields the dressing of the Higgs fields can be rather large
($>> 10 $). We  already calculated\cite{nmsgut3} the  dressing for
the \textbf{10}-plet component of the MSSM Higgs. However in our
realistic fits we find that the other components (in particular
those from the \textbf{210}) can form a significant fraction  of
the MSSM Higgs. Above  we showed  that a completely independent
line of argument \emph{requires} that the doublet $H$ be 50\%
derived from the \textbf{210}-plet ! Thus the lengthy
  calculation of the wave function corrections for each of the
 six GUT doublets  contributing   to the MSSM doublet is
necessary. Even from the partial calculation\cite{nmsgut3} one can
see that the large value of the wave function dressing makes the
GUT tree level matter fermion yukawa couplings (i.e
$\{h_{AB},g_{AB},f_{AB}\}_{tree}$ and therefore all the
$(y^f_{AB})_{tree}$ ) required to match the SIMSSM couplings at
$M_X^0$ much  smaller than they would be without these
corrections! It is important to note that this reduction in SO(10)
\textbf{16}-plet yukawa coupling magnitudes allows the $d=5$
baryon violation rates - which have always been problematically
large in supersymmetric GUTs - to be reduced to  acceptable levels
$\Gamma^{d=5}_{\Delta B\neq 0}< 10^{-34} yr^{-1}$. The NMSGUT
offers a novel and structural resolution of this longstanding
problem by taking seriously the non trivial wave function
renormalization of the light Higgs doublets of the MSSM by the
huge number of heavy fields they are coupled to. Since it is the
tree level couplings that enter the formulae for the inflaton
dynamics in the full GUT it is easier to satisfy
eqn.(\ref{hDeltavsM}). Because of this and the relatively large
value of  $M\sim M_3$ it should be be possible to achieve the
required fine tuning once the full wave function dressing is
computed.

The embedding in the GUT has overturned our naive assumption  that
the lowest intermediate scale would govern inflation. Instead it
is rather the largest. While setting us the  problem of finding
solutions to the tuning condition, compatible both  with an
accurate fit of fermion masses and acceptable values of
inflationary power spectrum and spectral index, it emphatically
shows that the soft terms have little role to play in the fine
tuning which belongs rather to the GUT and intermediate scale
physics only. Thus the physics of SIMSSM driven inflation is in
sharp contrast to the Dirac neutrino mass MSSM driven
inflation\cite{akm,hotchmaz} and makes it clear that they lie
counterpoised  not only as regards the nature of neutrino mass but
also as regards the nature of inflation and its regulating mass
scale besides their degree of naturalness. Note that the quadratic
dependence of corrections to soft susy parameters on the heavy
masses  as opposed to the logarithmic wave function normalization
of superpotential parameters makes the weaker fine tuning demands
on superpotential parameters only in the SIMSSM case even more
appealing.

   In Table I we give an example of the relevant parameters from  an accurate fit  of
the complete fermion spectrum in the NMSGUT : which has also been
tuned to make it as compatible as possible   with the inflationary
scenario presented here. The complete details regarding the fit
are given as Appendix I. It is apparent that the fine tuning
between the yukawas proceeds as anticipated with $1-\Gamma
=1=\Lambda_{n,d}$. The main problem lies in the fact that
$h^2/M\sim 10^{-19}$ GeV is too large by six orders of magnitude.
As a result the number of e-folds $N_{CMB}$  is much   smaller
than required. However as explained the formulae  used seriously
underestimate the Higgs wave function corrections. Moreover the
  search of the huge parameter space has just begun. Thus we are
  confident that this problem can also be overcome and a
  completely realistic fit compatible with inflation achieved.

Finally we remark that the single stage breaking of the simple
group $Spin(10)$ to the SM gauge group will lead to the  formation
of monopoles with a Kibble density  $n_K\sim M_X^3$ at the time of
the GUT phase transition.  However inflation by 50 or more e-folds
occurring  long after the epoch when the SO(10) monopoles are
formed will   dilute the monopoles  to completely levels removing
any monopole problem or signal.
 \begin{table}
 $$
 \begin{array}{|c|c|c|c|}
 \hline
 {\rm  Parameter }&Value &Parameter  &Value\\
 \hline
 \chi_{X}&        0.4458   & M_{h^0}&  122.99 \\
 \chi_{Z}&         0.1426 &   M_X & 7.08 \times 10^{17}\\
     f_3  &  1.066 \times {10}^{-3}& f_1 ,f_2 &2.59\times{10}^{-8},4.405 \times 10^{-5}   \\
     h & 2.44 \times 10^{-4} & \Lambda_{n} &0.999999\\
     M & 3.043 \times 10^{11}& \Lambda_{d} & 0.999999\\
     \Gamma & 4.343 \times 10^{-5}& \Delta_{tuning}& 0.989\\
     |{\overline{\sigma}}| & 4.69 \times 10^{15} &M_X &5.25 \times 10^{17}\\
A_0(M_X),m_0(M_X) & -5.235\times 10^5 , 1.260 \times 10^4 &\mu,B(M_X)  & 4.316\times 10^5,-1.128 \times 10^{11}\\
M^2_{\bar H} &   -1.498 \times 10^{11} & M^2_H & -1.448 \times 10^{11}\\
|\Delta_{H_0}|,|\Delta_{\bar{H}_0}| & 50.254,63.930 & |\alpha_{4}| & 0.707\\
M^{\nu^c}_3 & 4.86  \times 10^{13}&M^{\nu^c}_{1,2}  & 1.181 \times 10^{9},2.01\times 10^{12}  \\
 |y_{31tree}^\nu|& 1.997 \times 10^{-4}& |y_{21tree}^\nu|,|y_{11tree}^\nu| & 4.489 \times 10^{-5},1.640 \times 10^{-6}\\
 Log_{10}(h^2/M) & -18.706 & V_0,\phi_{end} & 3.579 \times 10^{52},2.153 \times 10^{15}\\
 N_{pivot},N_{CMB} & 54.22, 4.78 \times 10^{-4}  & \Delta , \beta & 8.82 \times 10^{-12},5.92 \times10^{-6}  \\
      \hline
 \end{array}
 $$
 \label{table I}\caption{\small{Illustrative example  of relevant parameters from an accurate fit
 of the fermion spectrum in the NMSGUT which is compatible with inflationary
  scenario. All masses are in GeV.
  $\chi_{X,Z}$ are the accuracies of the fits to 18 known fermion mass/mixing  parameters
   at $M_{X,Z}$.}}
 \end{table}

\section{Discussion}
 In this paper we have shown  how Supersymmetric Type-I seesaw models with
  the typical superpotential couplings found in MSLRMs and MSGUTs allow an attractive
   and natural implementation of renormalizable inflection point inflation.
    Inflation parameters are tied to seesaw parameter values and the required
     fine tuning is less severe and more stable than in the Dirac neutrino case since it is essentially
     independent of the supersymmetry breaking  parameters and is governed by the physics of
     intermediate scales $\sim 10^{6}- 10^{12} $GeV. In the Dirac
     neutrino case \cite{akm} the opposite it is true and the
     inflation occurs at low scales.

The post-inflationary reheating behaviour in the our  model
   differs  from the Dirac neutrino case. The mechanism of ``instant preheating''\cite{feldkoflinde} applied
   to inflection point inflation models shows that oscillation after  slow roll of a Susy flat direction
   inflaton    \cite{alfermaz}  ensures efficient transfer of all
   the inflaton energy into thermalized MSSM plasma within few
   Hubble times after the end of inflation and consequently a high
   reheat temperature $T_{rh}\sim 10^{11} - 10^{15} $ GeV.
   Thus this type of model requires a gravitino mass larger than
   about $50\, TeV $ to remain consistent with Nucleosynthesis. Such
  large Supersymmetry breaking scales are  also \emph{required} by the
  NMSGUT to fit all the fermion data\cite{NMSGUTs}. The high reheat
  temperatures and the presence of the Higgs in the inflaton
  sit comfortably with the requirements of
  thermal\cite{elliyana} and non thermal Leptogenesis\cite{ahnkolb}.
    The current work therefore extends the already wide
  scope of  the New Minimal Supersymmetric GUT  from  a completely
  realistic theory compatible with the  central paradigms of
  Beyond Standard Model(BSM) physics and predictive of parameters crucial to the
  discovery  of Supersymmetry. It has been shown to  potentially
  harbour a   consistent    Inflationary   cosmogony tied to the central paradigms of seesaw
  neutrino mass  and Leptogenesis. The complete calculation\cite{csaigck}  of the
  wavefunction  corrections to the tree level relations between
  SIMSSM and NMSGUT yukawa couplings will permit us to confirm the
  viability of our scenario in the NMSGUT context.

\section*{Acknowledgments}
 \vspace{ .5 true cm}
   We are grateful to Anupam Mazumdar and  Ling Fei Wang
 for correspondence and collaboration in earlier stages of this  work.
 C.S.A thanks David Lyth for discussions and useful comments.

 \eject

\begin{table}
{\bf{\large {Appendix }}}
 $$
 \begin{array}{|c|c|c|c|}
 \hline
 {\rm Parameter }&Value &{\rm  Field }&\hspace{10mm} Masses\\
 &&{\rm}[SU(3),SU(2),Y]&\hspace{10mm}( Units\,\,of 10^{16} Gev)\\ \hline
       \chi_{X}&  0.4458           &A[1,1,4]&   1323.58 \\ \chi_{Z}&
    0.1426
                &B[6,2,{5/3}]&            0.1912\\
           f_{11}/10^{-9}& 25.8969         &C[8,2,1]&{     60.14,    718.45,    741.09 }\\
           f_{22}/10^{-7}&440.5316    &D[3,2,{7/ 3}]&{     62.83,    715.80,    752.16 }\\
           f_{33}/10^{-2}&  0.1066     &E[3,2,{1/3}]&{      0.30,     46.93,     55.30 }\\
 h_{11}/10^{-6}&
 -0.8871+  0.1599i
                      &&{    55.303,    830.73,    942.57 }\\
 h_{12}/10^{-6}&
 -9.0533+  8.7699i
          &F[1,1,2]&     12.61,     12.61
 \\h_{13}/10^{-5}&
  5.8257+  0.4100i
                  &&     44.84,    664.62  \\
 h_{22}/10^{-5}&
 -4.1998+ 19.6580i
              &G[1,1,0]&{     0.048,      0.38,      0.75 }\\
 h_{23}/10^{-4}&
  8.3003-  4.1255i
                      &&{     0.755,     32.31,     32.44 }\\
 h_{33}/10^{-3}&
 -3.7417-  1.6595i
              &h[1,2,1]&{     1.176,     40.90,     62.13 }\\
 g_{12}/10^{-4}&
 -0.1179+  0.0940i
                 &&{   1120.35,   1178.92 }\\
 g_{13}/10^{-5}&
  5.9911-  0.5095i
     &I[3,1,{10/3}]&      0.67\\
 g_{23}/10^{-4}&
  9.1454-  9.2551i
          &J[3,1,{4/3}]&{     0.737,     29.75,     29.75 }\\
 \lambda/10^{-2}&
 -0.2982-  0.3350i
                 &&{     86.85,    798.73 }\\
 \eta&
-10.1628+  3.9777i
   &K[3,1, {8/ 3}]&{    100.76,    972.38 }\\
 \rho&
  0.4475-  2.1204i
    &L[6,1,{2/ 3}]&{     48.91,   1571.17 }\\
 k&
  0.0247-  0.0765i
     &M[6,1,{8/ 3}]&   1590.77\\
 \zeta&
  1.2522+  0.4940i
     &N[6,1,{4/ 3}]&   1582.18\\
 \bar\zeta &
  0.8170+  0.8221i
          &O[1,3,2]&   3043.86\\
       m/10^{16}GeV&    0.02    &P[3,3,{2/ 3}]&{     21.71,   2384.12 }\\
     m_\Theta/10^{16}GeV& -41.889e^{-iArg(\lambda)}     &Q[8,3,0]&     0.559\\
             \gamma&    3.78        &R[8,1, 0]&{      0.21,      0.82 }\\
              \bar\gamma& -3.5398     &S[1,3,0]&    0.9277\\
 x&
  0.9382+  0.6473i
         &t[3,1,{2/ 3}]&{      0.60,     38.05,     94.02,    181.09        }\\\Delta_X&      1.52 &&{    555.93,    755.27,  15333.10 }\\
              \Delta_{G}&  -7.505           &U[3,3,{4/3}]&     0.786\\
 \Delta\alpha_{3}(M_{Z})&  -0.004               &V[1,2,3]&     0.549\\
    \{M^{\nu^c}/10^{12}GeV\}&{0.001181,    2.01,   48.59    }&W[6,3,{2/ 3}]&           1877.78  \\
 \{M^{\nu}_{ II}/10^{-12}eV\}&  0.3880,    660.09,          15968.49               &X[3,2,{5/ 3}]&     0.185,    59.281,    59.281\\
                  M_\nu(meV)&{2.148903,    7.32,   40.17    }&Y[6,2, {1/3}]&              0.23  \\
  \{\rm{Evals[f]}\}/ 10^{-6}&{0.025897,   44.05, 1065.71         }&Z[8,1,2]&              0.81  \\
 \hline\hline
 \mbox{Soft parameters}&{\rm m_{\frac{1}{2}}}=
         -1062.672
 &{\rm m_{0}}=
         12603.819
 &{\rm A_{0}}=
         -5.2347 \times 10^{   5}
 \\
 \mbox{at $M_{X}$}&\mu=
          4.3160 \times 10^{   5}
 &{\rm B}=
         -1.1281 \times 10^{  11}
  &{\rm tan{\beta}}=           50.0000\\
 &{\rm M^2_{\bar H}}=
         -1.4978 \times 10^{  11}
 &{\rm M^2_{  H} }=
         -1.4480 \times 10^{  11}
 &
 {\rm R_{\frac{b\tau}{s\mu}}}=
  1.4504
  \\
 Max(|L_{ABCD}|,|R_{ABCD}|)&
          9.6779 \times 10^{ -23}
  {\,\rm{GeV^{-1}}}&& \\
 \hline\end{array}
 $$
 \label{table a} \caption{\small{Fit : Column 1 contains values   of the NMSGUT-SUGRY-NUHM  parameters at $M_X$
  derived from an  accurate fit to all 18 fermion data and compatible with RG constraints.
 Unificaton parameters and mass spectrum of superheavy and superlight fields are  also given.
 The values of $\mu(M_X),B(M_X)$ are determined by RG evolution from $M_Z$ to $M_X$
 of the values determined by the EWRSB conditions.}}\end{table}

\clearpage
 \begin{table}
 $$
 \begin{array}{|c|c|c|c|c|}
 \hline
 &&&&\\
 {\rm  Parameter }&Target =\bar O_i &Uncert.= \delta_i    &Achieved= O_i &Pull =(O_i-\bar O_i)/\delta_i\\
 \hline
    y_u/10^{-6}&  2.031523&  0.776042&  2.036884&  0.006908\\
    y_c/10^{-3}&  0.990278&  0.163396&  0.985427& -0.029685\\
            y_t&  0.350699&  0.014028&  0.350375& -0.023035\\
    y_d/10^{-5}&  7.314770&  4.264511&  8.249786&  0.219255\\
    y_s/10^{-3}&  1.385711&  0.654056&  1.335549& -0.076693\\
            y_b&  0.438505&  0.227584&  0.496225&  0.253620\\
    y_e/10^{-4}&  1.190847&  0.178627&  1.182832& -0.044871\\
  y_\mu/10^{-2}&  2.444540&  0.366681&  2.408165& -0.099201\\
         y_\tau&  0.519320&  0.098671&  0.529217&  0.100302\\
             \sin\theta^q_{12}&    0.2210&  0.001600&    0.2210&           -0.0066\\
     \sin\theta^q_{13}/10^{-4}&   29.4299&  5.000000&   29.5102&            0.0161\\
     \sin\theta^q_{23}/10^{-3}&   34.6272&  1.300000&   34.6440&            0.0129\\
                      \delta^q&   60.0211& 14.000000&   59.9431&           -0.0056\\
    (m^2_{12})/10^{-5}(eV)^{2}&    4.8973&  0.519109&    4.8979&            0.0012\\
    (m^2_{23})/10^{-3}(eV)^{2}&    1.5613&  0.312270&    1.5600&           -0.0043\\
           \sin^2\theta^L_{12}&    0.2939&  0.058780&    0.2944&            0.0094\\
           \sin^2\theta^L_{23}&    0.4618&  0.138552&    0.4597&           -0.0151\\
           \sin^2\theta^L_{13}&    0.0252&  0.019000&    0.0225&           -0.1439\\
 \hline
  Eigenvalues(\Delta_{\bar u})&   0.066017&   0.066029&   0.066044&\\
  Eigenvalues(\Delta_{\bar d})&   0.063539&   0.063551&   0.063566&\\
Eigenvalues(\Delta_{\bar \nu})&   0.073037&   0.073049&   0.073064&\\
  Eigenvalues(\Delta_{\bar e})&   0.080472&   0.080484&   0.080499&\\
       Eigenvalues(\Delta_{Q})&   0.061610&   0.061622&   0.061635&\\
       Eigenvalues(\Delta_{L})&   0.073586&   0.073599&   0.073611&\\
    \Delta_{\bar H},\Delta_{H}&       63.930186   &       50.254471    &{}&\\
 \hline
 \alpha_1 &
  0.6402+  0.0000i
 & {\bar \alpha}_1 &
  0.7220-  0.0000i
 &\\
 \alpha_2&
  0.0518+  0.0217i
 & {\bar \alpha}_2 &
  0.0387+  0.0540i
 &\\
 \alpha_3 &
 -0.0405-  0.0412i
 & {\bar \alpha}_3 &
 -0.0619-  0.0274i
 &\\
 \alpha_4 &
 -0.6968+  0.1200i
 & {\bar \alpha}_4 &
  0.6213-  0.0161i
 &\\
 \alpha_5 &
  0.1061+  0.0735i
 & {\bar \alpha}_5 &
  0.0585+  0.0173i
 &\\
 \alpha_6 &
  0.1356-  0.2204i
 & {\bar \alpha}_6 &
  0.1646-  0.2294i
 &\\
  \hline
                  |\alpha_1|, |\alpha_2| &0.640,0.056
         & |{\bar \alpha}_1 |,|{\bar \alpha}_2|&0.722,0.066
 &\\
                  |\alpha_3|,   |\alpha_4 |& 0.058,0.707
         & |{\bar \alpha}_3|,|{\bar \alpha}_4| &0.068,0.622
 &\\
                  |\alpha_5 |,  |\alpha_6| &0.129,0.259
         & |{\bar \alpha}_5| ,|{\bar \alpha}_6|&0.061,0.282
 &\\
 \hline
 \end{array}
 $$
 \label{table b} \caption{\small{Fit   with $\chi_X=\sqrt{ \sum_{i=1}^{17}
 (O_i-\bar O_i)^2/\delta_i^2}=
    0.4458
 $. Target values,  at $M_X$ of the fermion yukawa
 couplings and mixing parameters, together with the estimated uncertainties, achieved values and pulls.
 The eigenvalues of the wavefunction renormalization increment  matrices $\Delta_i$ for fermion lines and
 the factors for Higgs lines are given, assuming the external Higgs is 10-plet dominated.
 The Higgs fractions $\alpha_i,{\bar{\alpha_i}}$ which control the MSSM fermion yukawa couplings  are also
 given. Right handed neutrino threshold  effects   have been ignored.
  We have truncated numbers for display although all calculations are done at double
 precision.}}
 \end{table}

\clearpage
 \begin{table}
 $$
 \begin{array}{|c|c|c|c|}
 \hline &&&\\ {\rm  Parameter }&SM(M_Z) & m^{GUT}(M_Z) & m^{MSSM}=(m+\Delta m)^{GUT}(M_Z) \\
 \hline
    m_d/10^{-3}&   2.90000&   0.75028&   3.17037\\
    m_s/10^{-3}&  55.00000&  12.14614&  51.36207\\
            m_b&   2.90000&   3.19379&   3.05311\\
    m_e/10^{-3}&   0.48657&   0.45539&   0.46641\\
         m_\mu &   0.10272&   0.09267&   0.09776\\
         m_\tau&   1.74624&   1.71722&   1.71666\\
    m_u/10^{-3}&   1.27000&   1.10756&   1.28852\\
            m_c&   0.61900&   0.53583&   0.62338\\
            m_t& 172.50000& 143.44513& 171.73898\\
 \hline
 \end{array}
 $$
 \label{table c}
 \caption{\small{Values of standard model
 fermion masses in GeV at $M_Z$ compared with the masses obtained from
 values of GUT derived  yukawa couplings  run down from $M_X^0$ to
 $M_Z$  both before and after threshold corrections.
  Fit with $\chi_Z=\sqrt{ \sum_{i=1}^{9} (m_i^{MSSM}- m_i^{SM})^2/ (m_i^{MSSM})^2} =
0.1408$.}}

 $$
 \begin{array}{|c|c|c|c|}
 \hline
 {\rm  Parameter}  &Value&  {\rm  Parameter}&Value \\
 \hline
                       M_{1}&            276.93&   M_{{\tilde {\bar {u}}_1}}&          15629.87\\
                       M_{2}&            942.04&   M_{{\tilde {\bar {u}}_2}}&          15625.51\\
                       M_{3}&            662.69&   M_{{\tilde {\bar {u}}_3}}&          76919.36\\
     M_{{\tilde {\bar l}_1}}&           3892.50&               A^{0(l)}_{11}&        -323296.22\\
     M_{{\tilde {\bar l}_2}}&            283.29&               A^{0(l)}_{22}&        -322948.67\\
     M_{{\tilde {\bar l}_3}}&          65951.57&               A^{0(l)}_{33}&        -204476.65\\
        M_{{\tilde {L}_{1}}}&          23375.27&               A^{0(u)}_{11}&        -391038.17\\
        M_{{\tilde {L}_{2}}}&          23214.39&               A^{0(u)}_{22}&        -391035.62\\
        M_{{\tilde {L}_{3}}}&          52427.68&               A^{0(u)}_{33}&        -211710.78\\
     M_{{\tilde {\bar d}_1}}&           3610.56&               A^{0(d)}_{11}&        -322645.16\\
     M_{{\tilde {\bar d}_2}}&           3604.89&               A^{0(d)}_{22}&        -322642.31\\
     M_{{\tilde {\bar d}_3}}&         134282.49&               A^{0(d)}_{33}&        -125043.45\\
          M_{{\tilde {Q}_1}}&          17575.24&                   \tan\beta&             50.00\\
          M_{{\tilde {Q}_2}}&          17572.75&                    \mu(M_Z)&         351033.09\\
          M_{{\tilde {Q}_3}}&         109825.82&                      B(M_Z)&
          2.4726 \times 10^{  10}
 \\
 M_{\bar {H}}^2&
         -1.1964 \times 10^{  11}
 &M_{H}^2&
         -1.3584 \times 10^{  11}
 \\
 \hline
 \end{array}
 $$
 \label{table d} \caption{ \small {Values (GeV) in  of the soft Susy parameters  at $M_Z$
 (evolved from the soft SUGRY-NUHM parameters at $M_X$).
 The  values of soft Susy parameters  at $M_Z$
 determine the Susy threshold corrections to the fermion yukawas.
 The matching of run down fermion yukawas in the MSSM to the SM   parameters
 determines  soft SUGRY parameters at $M_X$. Note the  heavier third
 sgeneration.  The values of $\mu(M_Z)$ and the corresponding soft
 susy parameter $B(M_Z)=m_A^2 {\sin 2 \beta }/2$ are determined by
 imposing electroweak symmetry breaking conditions. $m_A$ is the
 mass of the CP odd scalar in the in the Doublet Higgs. The sign of
 $\mu$ is assumed positive. }}
 \end{table}

 \clearpage
 \begin{table}
 $$
 \begin{array}{|c|c|}
 \hline {\mbox {Field } }&Mass(GeV)\\
 \hline
                M_{\tilde{G}}&            662.69\\
               M_{\chi^{\pm}}&            942.04,         351033.11\\
       M_{\chi^{0}}&            276.93,            942.04,         351033.10    ,         351033.11\\
              M_{\tilde{\nu}}&         23375.180,         23214.295,         52427.637\\
                M_{\tilde{e}}&           3892.76,          23375.33,            277.93   ,          23214.55,          52422.21,          65955.95  \\
                M_{\tilde{u}}&          15629.83,          17575.16,          15625.45   ,          17572.68,          76918.45,         109826.62  \\
                M_{\tilde{d}}&           3610.66,          17575.35,           3604.97   ,          17572.86,         109823.42,         134284.46  \\
                        M_{A}&        1112118.78\\
                  M_{H^{\pm}}&        1112118.78\\
                    M_{H^{0}}&        1112118.78\\
                    M_{h^{0}}&            122.98\\
 \hline
 \end{array}
 $$
 \label{table e}\caption{\small{Spectra of supersymmetric partners calculated ignoring generation mixing effects.
  Inclusion of such effects   changes the spectra only marginally. Due to   large
 values of $\mu>>M_Z,M_W$ the LSP and light chargino are  essentially pure Bino and Wino($\tilde W_\pm $).
   The light  gauginos and  light Higgs  $h^0$, are accompanied by a light smuon and  sometimes  selectron.
 The rest of the sfermions have multi-TeV masses. The mini-split supersymmetry spectrum and
 large $\mu,A_0$ parameters help avoid problems with Flavor Changing Neutral Currents and Charge
 and Color breaking/Unbounded from below(CCB/UFB) instability\cite{nmsgut3}.
 The sfermion masses  are ordered by generation not magnituide. This is useful in identifying the spectrum
  calculated including generation mixing effects. Note the very
  light(right)  smuon.
  }}\end{table}
 \begin{table}
 $$
 \begin{array}{|c|c|}
 \hline {\mbox {Field } }&Mass(GeV)\\
 \hline
                M_{\tilde{G}}&            663.15\\
               M_{\chi^{\pm}}&            942.22,         351025.61\\
       M_{\chi^{0}}&            276.99,            942.22,         351025.60    ,         351025.60\\
              M_{\tilde{\nu}}&          23214.64,          23375.50,         52426.007\\
                M_{\tilde{e}}&            249.75,           3890.86,          23214.90   ,          23375.64,          52420.65,          65953.06  \\
                M_{\tilde{u}}&          15626.70,          15631.41,          17574.07   ,          17576.34,          76909.50,         109817.78  \\
                M_{\tilde{d}}&           3604.73,           3615.05,          17574.26   ,          17576.53,         109815.13,         134273.86  \\
                        M_{A}&        1112398.16\\
                  M_{H^{\pm}}&        1112398.16\\
                    M_{H^{0}}&        1112398.15\\
                    M_{h^{0}}&            122.99\\
 \hline
 \end{array}
 $$
 \label{table f}\caption{\small{Spectra of supersymmetric partners calculated including  generation mixing effects.
 Inclusion of such effects   changes the spectra only marginally. Due to   large
 values of $\mu>>M_Z,M_W$  the LSP and light chargino are  essentially pure Bino and Wino($\tilde W_\pm $).
  Note that the ordering of the eigenvalues in this table follows their magnitudes, comparison
  with the previous table is necessary to identify the sfermions}}\end{table}

\end{document}